\documentclass[12pt,preprint]{aastex}

\slugcomment{Submitted to ApJ}
\shortauthors{Howell et al.}
\shorttitle{Fun with EF Eri}

\begin{document}

%
%

\title{Mass Determination and Detection of the Onset of Chromospheric Activity 
for the Sub-Stellar Object in EF Eridani}

\author{Steve B. Howell\altaffilmark{1}, 
Frederick M. Walter\altaffilmark{2}, 
Thomas E. Harrison\altaffilmark{3}, \\
Mark E. Huber\altaffilmark{4},
Robert H. Becker\altaffilmark{4,5},
Richard L. White\altaffilmark{6}
}

\altaffiltext{1}{WIYN Observatory and National Optical Astronomy Observatory,
950 N. Cherry Ave, Tucson, AZ 85719 {\it howell@noao.edu}}
\altaffiltext{2}{Dept. of Physics and Astronomy, Stony Brook University
{\it fwalter@astro.sunysb.edu}}
\altaffiltext{3}{New Mexico State University
{\it tharriso@nmsu.edu}}
\altaffiltext{4}{IGPP/Lawrence Livermore National Laboratory, Livermore, CA 94550
{\it mhuber@igpp.ucllnl.org}}
\altaffiltext{5}{Dept. of Physics, University of California at Davis, Davis, CA 95616}
\altaffiltext{6}{Space Telescope Science Institute, 3700 San Martin Dr,, Baltimore, MD
21218}

\begin{abstract}
EF Eri is a magnetic cataclysmic variable that has been in a low 
accretion state for the past nine years.
Low state optical spectra reveal the underlying Zeeman-split
white dwarf absorption lines. These features 
are used to determine a value of 13-14 MG as the white dwarf field
strength.
Recently, 5-7 years into the low state, Balmer and other emission
lines have appeared in the optical. 
An analysis of the H$\alpha$ emission line yields the first
radial velocity solution for EF Eri, leading to 
a spectroscopic ephemeris for the binary and, using the best available
white dwarf
mass of 0.6M${\odot}$, a mass estimate for the secondary 
of 0.055M${\odot}$.
For a white dwarf mass of 0.95M${\odot}$, the average for magnetic
white dwarfs, the secondary mass increases to 0.087M${\odot}$.
At EF Eri's orbital period of 81 minutes,
this higher mass secondary could not be a normal star and still 
fit within the Roche lobe.
The source of the Balmer and other emission lines is confirmed to be 
from the sub-stellar secondary and we argue that it is due to stellar activity.
We compare EF Eri's emission line spectrum and activity behavior to that recently
observed in AM Her and VV Pup and attributed to stellar activity.
We explore observations and models originally developed for V471 Tau, 
for the RS CVn binaries, and for extra-solar planets.
We conclude that irradiation of the secondary in EF Eri and similar 
systems is unlikely and, in polars,
the magnetic field interaction between the two stars (with a possible tidal
component) is a probable mechanism which would 
concentrate chromospheric activity on the secondary 
near the sub-stellar point of the white dwarf.
\end{abstract}

\keywords{stars: activity --- stars: low-mass, brown dwarfs --- stars: individual (EF Eri, AM Her, VV Pup)}

\section{Introduction}
EF Eridani has become a binary star of renewed interest in recent years. This is primarily due to the fact that 
the mass accretion from the low-mass secondary to the highly magnetic white dwarf has been essentially stopped
for the past 9 years. During this period of very low accretion, observers have flocked to telescopes around the world
in hope of detecting and understanding the component stars that make up EF Eri. As we will see, it has taken 
a small 1.5-m telescope, part of the SMARTS consortium, to provide the first 
new insights into the true nature of this fascinating (and confounding) binary.

EF Eri was discovered over 30 years ago as a hard X-ray source 
(2A 0311 -227; Cooke et al. 1978). It was quickly identified optically as a 
14th magnitude blue source (Griffiths et al. 1979)
with properties similar to AM Herculis (Charles and Mason 1979). AM Her, 
the not-so-typical prototype of the AM Her class of
interacting binaries, was the first member of a new class 
of cataclysmic variable (Tapia 1977). Today, these magnetic cataclysmic 
variable (CV) binary systems are called polars (due to their polarized light),
systems in which matter is accreted onto the more massive primary
from the lower mass secondary. The primary star is a highly magnetic white 
dwarf (B$\sim$10 MG to 250 MG) and the mass donor ranges
from normal M stars (as in AM Her) to sub-stellar brown dwarf-like 
objects which are believed to reside in EF Eri and other ultra-short period CVs
(see Warner 1995 for a review).
 
Polars show large, random changes in their brightnesses on weeks to months,
and even year timescales.  In the high state, when mass accretion is in full 
swing, their luminosities are dominated by an accretion-produced continuum 
from the X-ray to the IR, superposed with strong emission lines from hydrogen
and helium. When the mass 
transfer slows or stops, the underlying stars are revealed, and these dominate
the optical and infrared spectral energy distribution. The origin for
the starting and stopping of the mass transfer from the low mass companion is 
not understood, but stellar activity cycles in the synchronously locked, 
rapidly spinning late type secondary has been a long standing 
explanation (see Bianchini, 1992; King and Cannizzo 1998; Hessman et al. 2000) 
which has recently gained some observational support (Kafka et al. 2005, 2006; 
Mason et al. 2006).

A number of recent papers have been dedicated to EF Eri 
(e.g., Beuermann et al. 2000; Harrison et al. 2003, 2004) and they track
the evolution of this system into its low state, where the system evolved to
the point where the optical spectrum was emission-line free, and dominated by 
a cool white dwarf that showed Zeeman-split Balmer absorption lines (e.g., see 
Fig. 2 in Harrison et al. 2004). The secondary in EF Eri has not yet
been conclusively identified, but both Beuermann et al. (2000), and Harrison
et al. (2004) suggest that it must be brown dwarf-like object, and recent
$Spitzer$ observations detect strong mid-infrared emission that suggest 
this object is an L or T dwarf (Howell et al., 2006).

We have been monitoring EF Eri with regularity over the past 9 years, using its
continued low state as an invitation from the binary to attempt to learn about
its stellar components. During the Fall of 2004, we noticed that EF Eri, while 
becoming no brighter, began to show weak, narrow H$\alpha$ emission. The 
remainder of the Balmer series, and other spectral lines, are also now present.
Detailed analysis of the H$\alpha$ emission line yields the first dynamical 
orbital velocity solution for EF Eri. We use this information and the remaining 
spectral features to determine the source of the emission and show that it is 
consistent with chromospheric activity on the sub-stellar secondary.

\section{Observations}

We present a number of observations in this paper from a variety of sources. 
The largest data set is from our monitoring program of EF Eri using the SMARTS 
telescopes at CTIO. These observations consist of multi-color photometry
and optical spectroscopy covering the H$\alpha$ region. 
We also present a single blue spectrum obtained in Fall 2005 at the Apache 
Point Observatory and use it to compare EF Eri to the prototype polar, AM Her. 
The comparison reveals that both stars have white dwarfs with similar magnetic 
field strengths and both stars show stellar activity indicators during their low 
state. Finally, we present a high S/N, low state  optical spectrum from Keck II 
revealing the full panorama of on-going spectral activity in EF Eri.

\subsection{SMARTS}

Photometric and spectroscopic monitoring of EF Eri has been accomplished using 
the SMARTS\footnote{
SMARTS, the Small and Medium Aperture Research Telescope Facility, is
a consortium of universities and research institutions that operate the
small telescopes at Cerro Tololo under contract with AURA.}
facilities at Cerro Tololo. This program has operated from Aug. 2003 to the 
present and all data were obtained by the SMARTS service observers.

\subsubsection{Photometry}

The optical images were obtained using the ANDICAM dual channel
imager on the SMARTS 1.3m telescope. 
ANDICAM obtains simultaneous optical and near-IR images, although 
EF Eri is too faint in its low state for us
to make use of the near-IR capability. The optical detector is a Fairchild 447 2048x2048
CCD. We obtained 100 second integrations through the B, V, and I filters. 
The images are processed (overscan subtraction and
flat-fielding) prior to distribution to SMARTS observers.
Because we obtained only single images through each filter each night,
the noise is dominated, in some cases, by cosmic rays and hot pixels.

We obtained 90 sets of useful
images between 22 August 2003 and 21 December 2005.
On 2 November 2005 UT we obtained continuous coverage of 1.09 orbital cycles
in 36 V-band images between 2.34 and 3.81 hours UT (see \S2.1.2).

We measured the star's brightness using differential photometry with a
4$\arcsec$ radius aperture.
The photometry is differential with respect to 5 nearby stars. 
One of these stars appears to a long term variable that has faded by
about 0.06 mag in 
$B$ and $V$, and 0.3~mag in $I$ over our 850 day baseline. This object has been removed
as a valid comparison star.
The internal consistency of the remaining four comparison stars is $<$5~mmag.
Uncertainties in the magnitudes of the target are dominated by counting statistics.

Figure 1a shows the SMARTS differential photometry in the $B$, $V$, and $I$ 
bands for EF Eri spanning the last few years. 
We can use the SMARTS comparison stars and our previous calibrated 
observations of the EF Eri field
(Harrison et al. 2003) to say that EF Eri has been amazingly constant in the
optical during this time period; EF Eri 
has remained in the low state with $V$$\sim$$B$$\sim$18.2 and $I$$\sim$17.8.
As an additional check on the state of EF Eri, we examined
new K band images on 2005 July 30 UT kindly taken at the
United Kingdom Infrared Telescope (UKIRT) and provided by S. Leggett and T. Geballe.
Upon reduction, these images showed EF Eri to have a $K$ magnitude of 15.2$\pm0.2$, the same 
value it has had for the past 8 years.

Figure 1b phases the Fig. 1a observations on the EF Eri
orbital period of 81 min and our new spectroscopic 
ephemeris discussed below. The sine-like
shape of the $B$ light curve, with semi-amplitude near 0.1 mag, is similar
to, but less well defined than
in $V$ and appears to be absent altogether in $I$.
These light curve features show the same level, shape, and phasing as light
curves presented in Harrison et al. (2003; their Fig. 2 shows observations
obtained in 2001 Dec.
\footnote{Note, Harrison et al. phased their light curves on the
original Bailey photometric ephemeris.}). 
Similar to the Harrison et al. interpretation, we believe the weak blue
photometric modulation is due to a surface feature on the white dwarf. 
We will discuss below (\S3.3)
whether this feature is consistent with a bright, warm region or a cool, 
higher opacity
region near one of the accretion poles.

\subsubsection{Spectroscopy}
Our first two SMARTS spectra of EF Eri were obtained on 2004 Aug 13 UT and 
on 2004 Oct 19 UT. The two spectra look nearly identical and the 
second one is presented in Fig. 2.
These spectra were obtained with the 1.5m SMARTS telescope using 
grating 13 providing 17\AA~spectral resolution and covering
most of the optical (3200\AA-9500\AA). We note in Fig. 2 (as well as in the 13 Aug
observation) the clear signature of the first two Balmer lines in 
emission (H$\alpha$ eq. width is -19\AA\, see \S3.3),
He I emission at 5876\AA, and the probable detection of the underlying
Zeeman split absorption lines from the magnetic white dwarf. 

Given our detection of Balmer emission, we continued monitoring EF Eri 
ultimately obtaining 10 sets of spectra with the RC spectrograph
on the 1.5m SMARTS telescope starting on 20 Oct 2004 UT.
(see Table 1). The RC spectrograph is
a slit spectrograph with a 300\arcsec\
long slit oriented E-W and a Loral 1200x800 CCD detector. 
We made all our observations through a 110$\mu$m (1.5\arcsec) slit.
EF Eri has a nearby unrelated companion star approximately 22\arcsec\ west-southwest.
In order to orient the SMARTS observers with our field and to make completely 
sure we did not
intersperse or include spectra of this nearby object,
we purposefully obtained a spectrum of
this star. This companion 
star appears to be a weak-lined, late-type star ($\sim$K) and has absolutely
no spectral similarities to EF Eri: no emission lines, is significantly brighter\footnote{Note that in
the familiar finding chart for EF Eri available from the Downes CV Atlas at 
http://archive.stsci.edu/prepds/cvcat/ and other places, 
the nearby companion star and EF Eri appear similar in
brightness. This image was obtained at a time when EF Eri was in a high (bright) state near
$V$=14.}, and is easily avoided in the spectrograph slit.

We obtained at least three spectral images at each epoch in order to filter cosmic rays.
The basic spectral reductions subtract the overscan and divide each image by 
the normalized flat field image using a pipeline written in IDL.
The spectrum in Figure 2 has 17 \AA\ resolution and covers
     the entire optical spectrum out to 9300\AA. All subsequent spectra were
     obtained with grating 47
in first order, with a wavelength
range of 5652-6972\AA\ and a resolution of 3.1\AA. Wavelength calibration
utilizes a Neon lamp exposure, taken at the start of the observing
sequence.
Exposure times for the individual frames range from 500 to 1200 seconds.
We generated a median image from all the images at each of the 10 epochs.
As these longer exposure times caused the spectral signal to be
integrated over a significant fraction of the 81 minute orbital period, any 
radial velocity or line profile information, is smeared in the summed images.  

EF Eri is a challenging target for spectroscopic work at a 1.5-m telescope.
The SMARTS service observers are quite skilled and highly proficient and performed the
spectral observations with great care.
The SMARTS spectra of EF Eri on all occasions presents a weak (pseudo)continuum, 
with a typical S/N per pixel in the continuum $<$3. 
We extract the spectra from the images in two ways. First, we use a boxcar
extraction with an extraction width set by a fit to the cross section of the
spectrum at chip center. The background is measured above and below the spectrum,
and linearly interpolated to the position of the source spectrum. 
Secondly, we fit a Gaussian atop a flat background at each column in the
chip. In both cases, the location of the spectrum (the trace) is determined
from a bright star observed that night, and shifted to match the position of
the target. The width of the spectrum as a function of wavelength
is determined from the trace of the
bright star. The extraction width is not allowed to vary in the Gaussian fits.
The flux from the Gaussian extraction is the analytical integration of the 
Gaussian fit.

We observed a spectrophotometric standard, either Feige 110 or LTT 4364,
each night in order to convert the counts spectrum to a flux spectrum.
Due to seeing-related slit losses, we do not obtain absolute fluxes, but rather
use the standards to recover the shape of the continuum.

We know that the Balmer emission produced by the high state accretion in EF Eri
greatly declined after the start of its 1997 Jan. low state (Wheatley \& Ramsey 
1998; Beuermann et al. 2000). Spectra obtained in Nov. 2000 (Reinsch et al. 
2004; Euchner et al. 2002) show very weak Balmer emission with only H$\alpha$ 
being above the continuum. By March 2001, the Balmer emission had disappeared 
altogether (Harrison et al. 2003\footnote{In
Fig. 4 of Harrison et al., the authors claim that the H$\beta$ emission line 
has weakened. A reexamination of this spectrum with respect to the pure Zeeman
absorption spectrum observed in 2002 as well as model fits such as in Reinsch
et al. 2004, show that it is likely that no Balmer emission was present in the
Harrison et al. spectrum.}) and remained
absent as late as Aug. 2002 (Harrison et al. 2004; Howell 2004).
On 13 Aug 2004 UT, 
we noted the presence of Balmer and He emission in EF Eri 
and thus began our SMARTS monitoring efforts. 
Therefore, we can only state that the H I emission in EF Eri started sometime
between 2002 Aug. and 2004 Aug. 

While our spectra are too poor for continuum flux measurements, in nearly all 
of them, H$\alpha$ emission was clearly seen, and we could obtain a measurement 
of its line center and equivalent width. To accomplish this, we Fourier-smoothed 
the individual spectra (see Figure 4), and then fit the spectra between 6400 
and 6700\AA ~ with the sum of a quadratic background and a Gaussian line. The 
S/N of the (pseudo)continuum in our data is far too low to reveal 
any photospheric features from the white dwarf. The H$\alpha$ emission line is 
generally narrow, but can be significantly velocity-smeared even in a ten minute
integration. We use the best-fit position of the Gaussian for the radial
velocity analysis. A simple centroid of the emission line gives comparable 
results.  Figure 5 shows one of the trailed SMARTS spectra of the H$\alpha$ 
emission line in EF Eri. The
plot illustrates the obvious sinusoidal motion of the H$\alpha$ emission line 
as well as its modulation with phase.

\subsection{Apache Point Observatory}

Optical spectroscopy of EF Eri and AM Her were obtained on
2005 November 6 UT using the Double Imaging 
Spectrograph\footnote{http://www.apo.nmsu.edu/Instruments/DIS/Default.html}
on the Apache Point Observatory 3.5 m. We used the high resolution gratings
with a 1.5 arc-second slit to deliver a dispersion of 0.6 \AA/pix. The central 
wavelength of the blue spectrum was 4500 \AA, covering the spectral range 
$\lambda \lambda$3865 $-$ 5094 \AA. The exposure times for both AM Her and
EF Eri were 20 minutes each, this is one-fourth of the EF Eri orbital period.
The exposures were started at 6:21:56.0 11/06/2005 UT (EF Eri) and 
1:18:30.4 11/06/2005 UT (AM Her). Both AM Her and EF Eri were in 
a low state during this time (see Kafka et al. 2005, 2006). 

Figure 6 presents our APO observations of EF Eri and AM Her.
The spectrum of AM Her has been over-plotted at a flux level about ten times 
below its real value.  In Fig. 6, we note that the 
spectrum of AM Her rises to the blue, while EF Eri drops, a consequence of the 
difference of the white dwarf temperatures in the two polars.
Balmer, Ca II (H\&K), and He I emission is seen in both systems and 
and has been associated with the secondary star in AM Her. 

\subsection{Keck}

On January 03, 2006 UT a 10 minute, medium resolution spectrum of EF Eri
was obtained with ESI (Sheinis et al. 2002) in echellette
mode on the Keck II telescope.
This setup provided a useful spectral coverage of 4000--10,000\AA\ in
10 spectral orders with a constant dispersion of 11.4 km $s^{-1}$
pixel$^{-1}$.
A slit width of 1" aligned along the parallactic angle was used 
with a plate scale of 0.127--0.183" pixel$^{-1}$ providing a resolution of
3000 to 5000 across the optical spectrum, giving 1.6\AA\ resolution 
at H$\alpha$.
Wavelength calibrations were accomplished using Hg-Ne-Xe and Cu-Ar lamps.
The active flexure control on ESI minimizes drift in the wavelength
calibration and/or any fringing that is crucial for clean sky subtraction.
The spectrophotometric standard Feige 110 was used for flux calibration.
The data were reduced using standard IRAF routines with additional custom
analysis routines to properly correct for the large curvature,
strong nonlinear photometric response, and small tilt of sky lines.

Figure 7 shows our high S/N Keck spectrum of EF Eri. The underlying 
white dwarf continuum is now
revealed, presenting the Zeeman split Balmer absorptions from the white dwarf.
Figure 7 also shows the presence of narrow emission lines of 
the entire Balmer series of hydrogen, the Pickering series of He I 
(4388, 4471, 4922?, 5876, 6678, and 7065 \AA), Na I, and
the Ca II triplet (8498, 8542, 8662 \AA). No He II emission is observed.
The continuum slope change seen near 8600\AA,
also noted by Beuermann et al. (2000), is not a contribution from the 
secondary 
but, as shown by Harrison et al. (2003, 2004), maybe due to
a weak cyclotron hump feature consistent with 
the 13-14 MG magnetic field present in EF Eri (see \S3.2).
Our high S/N Keck spectrum and our low resolution SMARTS spectrum (Fig. 3)
confirms our belief that the SMARTS radial velocity data samples 
only the top of the H$\alpha$ emission line sitting on a pseudo-continuum.

\section{Analysis}

\subsection{Radial Velocity Solution}

The SMARTS spectral observations can be discussed as two groups.
The first group (group one) are data specifically obtained for radial velocity and orbital motion determination.
They all have a uniform integration time and were collected over a short time 
period in 2005 Nov, 2005 Dec (two separate nights), and 2006 Feb.
The remaining SMARTS spectra (group two) consist of our monitoring observations
obtained between 2004 Oct and 2005 Nov (See Table 1).
The four datasets that form group one cover at least one orbital period each 
in a continuous manner.
Group two consists of spectra usually taken in threes and obtained in the months of 
2004 Oct \& Nov and 2005 Jul, Sep, and Nov. These latter data have various
integration times, from  500 to 1200 seconds, with half being 1200 seconds or 
1/4 of the orbital period of EF Eri.

Bailey et al. (1982) obtained optical and near-IR high state photometry for EF 
Eri over the observing season in Fall 1979.
Those authors noted that the optical and near-IR light curves were complex and did not agree 
in shape over the various wavelengths, but the infrared light curves showed, on most nights, a
narrow dip feature that nicely repeated. The center of this narrow dip was used to set 
a photometric ephemeris for EF Eri whereby the dip was
chosen to be phase 0. Bailey et al. discuss the various light curves and conclude that 
the narrow dip was most likely caused by the secondary 
eclipsing the gas stream as it traveled toward the white dwarf. 
It was thus assumed that Bailey's photometric phase 0.0 was near the true spectroscopic phase 0.0.
All work on EF Eri since the Bailey et al. paper 
has used this photometric ephemeris and phasing as no other was available.

We initially focused on the group one data set, 45 spectra in total, and phased 
them on the orbital period given by Bailey et al. (1982). 
Our H$\alpha$ line center velocity measurements are
presented in Figure 8. Each continuous set of observations for the four nights are represented
by a different color as described in the figure caption.
While our formal error in the measurement of the H$\alpha$ line center is $\sim$1\AA\
for most of the measurements, the spectra containing weaker emission lines (near phase 0.0)
are a bit worse. Thus, in fitting our orbital solution, we used
conservative 1$\sigma$ error bars of $\pm$1.5\AA\ for each of the 
radial velocity measurements.
We then added in the group two SMARTS spectra and produced
a new radial velocity solution. The inclusion of the additional points 
produced a very similar (within the formal errors) 
result to that obtained with the group one data alone.
This consistent behavior providing confirmation
that the H$\alpha$ emission has remained constant in velocity behavior over the
entire time we have detected it. 
Figure 8 shows the additional velocity measurements as a separate color.
The Bailey et al. photometric phases are given on the top axis in Fig. 8
to illustrate the relationship between the 
older photometric phase and our new dynamic spectroscopic phasing. 

Our best fit sine curve solution to the
H$\alpha$ emission line centers for all our spectra
yields a normal CV spectroscopic {\it primary star red-to-blue} crossing ephemeris 
for EF Eri of
$$ T_0 = HJD 2453716.61108(5) + 0.05626586(80) E $$
where orbital phase 0.0 is the secondary's inferior conjunction.
We have assumed that the orbit is
circular and the components make their respective red-to-blue crossings 
180 degrees apart in phase and have used the
orbital period determined by Bailey et al. (1982).
Our best fit to the H$\alpha$ measurements yields a K$_2$ of 269 $\pm$ 18 
km sec$^{-1}$ and $\gamma$=115 $\pm$ 15 km sec$^{-1}$. 
Spectroscopic phase 0.0 (0.5) is equal to Bailey et al. photometric phase 0.41 (0.91).
Note that all other plots in the paper use 
our new spectroscopic orbital phasing derived above.

The large value for K$_2$ indicates that the emission is produced 
on, or near, the secondary given its large value. It would be impossible 
for the white dwarf in this binary to have such a large K amplitude.
We will show below that this emission, as appears true for both AM Her and 
VV Pup during their low states, is not caused by irradiation. Given our new 
dynamic orbital solution, the suggestion that the narrow photometric dip 
feature used by Bailey et al. to define phase 0.0 is caused by some sort of 
gas stream eclipse by the secondary cannot be correct.

\subsection{Zeeman Absorption Spectroscopy and Cyclotron Emission}

The nearly identical Zeeman absorption structures from the
white dwarfs in both AM Her and EF Eri are easily seen in Fig. 6.
AM Her has a listed magnetic field strength of 12.5 MG 
(Bonnet-Bidaud et al. 2000) 
while EF Eri's field strength has been estimated in various ways 
to be between 17-35 MG (Warner 1995; Reinsch et al. 2004). 
Using Fig. 2 in Harrison et al. (2004) and our Keck spectrum 
presented here (Fig. 7) we can measure the Zeeman split $\pm\sigma$ components 
in the H$\alpha$ line to determine the surface magnetic field strength in EF Eri. 
We find the $\pm\sigma$ components are split from the main component of H$\alpha$ by
28$\pm$1.4\AA\ in both the older spectrum and our new Keck observation. Using  
Eq. 10 in Wickramasinghe \& Ferrio (2000) for the linear regime, this splitting yields
a surface field strength of 13.8$\pm$1 MG for the white dwarf in EF Eri.
We can also match our Fig. 7 H$\beta$ profile in some detail to the model work
presented in Fig. 2 of Euchner et al. (2005). Doing so, yields a best 
magnetic field estimate of 13-14 MG. A rough comparison of our H$\alpha$ and H$\beta$ $\pm\sigma$ profiles
with the detailed modeling for single, highly magnetic white dwarfs presented in Euchner
et al. (2005, 2006) shows that we can estimate the linear Zeeman surface field 
to about $\pm$1 MG, but can do no better with these data. 

Using our new spectroscopic phasing for EF Eri, we note that the optical light 
curve (Fig. 2) shows an asymmetric hump-like modulation consistent 
with a slight 
brightening (0.1 mag) centered near phase 0.4. The amplitude of this hump
increases toward the blue. Phase 0.4 is where we would
expect to be viewing the magnetic white dwarf at near right angles to the
magnetic poles, thereby suggesting the slight flux increase may be related to
cyclotron emission.  Even though EF Eri is in a low accretion state, 
it is not zero as cyclotron humps, consistent with a 13-14 MG field, are
seen in infrared spectra (Harrison et al. 2005). 
The low state light curve could also be interpreted
as having a luminosity dip from 0.8 to 0.0. 
This phase is where we would expect to have a more or less direct view of the
main accretion pole on the white dwarf and/or the back end of the 
secondary . Given that the amplitude of the light curve increases
to the blue, a cooler secondary back end seems to be ruled out.
Perhaps we are seeing a dark spot on the white dwarf near the main accretion
pole due to a region of higher opacity caused by atmospheric 
metals migrating toward the magnetic pole an forming a metallic 
lake (as seen in the
single white dwarf GD394, Dupuis et al. 2000).
This idea, however, is difficult to reconcile given the high UV 
flux levels observed in EF Eri
in recent GALEX observations (Szkody et al., 2006).
Weak mass accretion appears to be a common behavior in low state polars (e.g., Pandel \& Cordova 2004; Howell et al. 2005)
and is likely due to connected magnetic field lines between the two stars
allowing the secondary wind (flares, prominences, etc.) 
to be accreted onto the white dwarf (see below).

\subsection{The Origin of the Optical Emission Lines}

It is useful to compare the current spectrum of EF Eri to the low state
spectra of AM Her and VV Pup. All three show emission from H I and Ca II.
While He I emission is clearly seen in VV Pup (Mason et al. 2006) and AM Her (Kafka et al. 2006), it is
only confidently detected in our Keck spectrum of EF Eri.
AM Her (P$_{orb}$=3 hr, M4V secondary) has been in a low state much 
of the past two years  and Kafka et al. (2005, 2006) attribute the low state Balmer emission 
observed in AM Her to stellar activity on the secondary star. Their claim is 
based on a detailed examination of the Balmer emission, particularly H$\alpha$, 
including its radial velocity amplitude, its narrowness, and the line strength 
behavior throughout the orbit. Their analysis eliminates irradiation as the
dominant source for the Balmer emission seen in AM Her during the low state. 
VV Pup, (P$_{orb}$=100 min) with an M7V secondary star, has also been in a low 
state for much of the past few years and the origin of its emission line
spectrum has also been attributed to stellar activity on the secondary star
 (Mason et al. 2006).
Both AM Her and VV Pup seem to show secondary star emission features 
at all times
during their recent low states. EF Eri's secondary, however, 
appears to not have had line emission for the first $\sim$5-7 years of its low
state. If the onset of the emission lines in EF Eri corresponds to the start of
a stellar activity cycle, as we believe, this event did not cause EF Eri to
immediately go into a high state. The activity in AM Her and VV Pup also seems 
to be uncorrelated with their high/low states starting or stopping. Thus, there 
does not appear to be a simple one-to-one correspondence between start/stop of 
stellar activity and high/low state transition in polars.

Our radial velocity curve makes it clear that the H$\alpha$ emission in
EF Eri comes from the secondary, but its source and the source of the Balmer, Pickering, 
Na I, and Ca II H\&K emission is an open question. Irradiation and stellar 
activity are the two obvious choices for the cause. 


\subsubsection {Irradiation or Stellar Activity?}

While irradiation is seen in some close interacting binaries, we know of no 
process whereby the cool white dwarf in EF Eri could suddenly start irradiating the 
secondary without any evidence for a change in the accretion rate.
We also note that AM Her and VV Pup in their recent low 
states show no indication of irradiation induced or enhanced Balmer or other 
emission, even given their hotter white dwarfs (see \S4).  However, Kafka et al. 
(2006) conclude that it is possible that some low level irradiation may be
present in AM Her during the low state but it is difficult to disentangle it 
from the much stronger activity-induced emission. 

The presence of Balmer, Na I, Ca II (H\&K and the IR triplet), and He I emission
are all the usual signatures of stellar activity 
(Giampapa et al. 1978) as these lines form in the high photosphere / low chromosphere
of the stellar atmosphere (see Vernazza et al. 1973).
It has always been suspected that the secondary stars in CVs should be highly active,
as they are rapidly rotating. However, there is little direct observational evidence 
of this assumption except in the
polars during low states (e.g., Schwope et al. 2001).

Figure 9 presents the equivalent width (EW) of
the H$\alpha$ line from our group one SMARTS spectra 
phased on the orbital period. We present the data for each night 
using a different color, 
the same scheme as in Fig. 8. If the H$\alpha$ line were produced
by irradiation, we would expect the EW to be a strongly peaked 
with a maximum near spectroscopic phase 0.5 
(the white dwarf facing side of the secondary), a result inconsistent with
the measurements.
What we see in Fig. 9 instead, is a nearly constant EW value, albeit
slightly variable, and a possibly reduced, but non-zero, 
line width near phase 0.0, the back side of the secondary.
As in AM Her and VV Pup, we find that the emission, and probably the stellar
activity processes, are generally concentrated toward the white dwarf 
facing side of the secondary. A stronger magnetic field on the white dwarf and/or
the alignment of active regions on the secondary star with the magnetic field 
(Simon et al. 1980) may cause
a stronger ``front side" concentration of activity. 
In addition, a tidal coupling
model presented in Rottler et al. (2002) as a cause of white dwarf facing
activity concentration will certainly be active in polars (and all CVs) and
may help, or be the primary cause, of a longitudinal enhancement.

The concentration of stellar activity and starspots on the secondary, 
at or near L1, has been previously assumed in the literature.
For example, Hessman et al. (2000) presented a model
to explain the high /low state behavior in AM Her. They used the modulated mass
transfer levels in this polar as an indicator of the amount of starspots needed
near L1. They concluded that nearly one-half of the ``L1 region" of the 
secondary star 
would have to be covered with spots during a high state. Furthermore, Hessman et
al. determined that this ``L1 area concentration" could only be possible 
if the spots were somehow
forced to wander into the region, possibly by a cyclic $\alpha^2$-$\Omega$
dynamo\footnote{Note that in CVs with orbital periods below the period gap, the
secondary stars are believed to have internal structures (fully convective 
or brown dwarf-like, e.g., VV Pup and EF Eri respectively) 
which are not expected to 
operate the $\alpha^2$-$\Omega$ dynamo in the same manner, or at all,
compared with CV secondary stars above the period gap (e.g., AM Her).}.
While this extreme level of starspot coverage has not been observed in AM
Her or other polars during a high state, it is easily hidden  by the
much higher accretion luminosity.


While the EW of the H$\alpha$ emission above the (pseudo)continuum is 
relatively constant over all phases for a given orbit of EF Eri, 
there is evidence for long term changes of up to a factor of 2. Figure 10 
presents the long term variations in the H$\alpha$ emission line
equivalent width.
The EW values have been integrated over each set of
observations, or at least half the orbital period, so should not be affected
by motions of the line across the underlying white dwarf Zeeman split 
absorption.  Assuming the white dwarf continuum is not
varying, and there is no substantial accretion, 
the observed variations in H$\alpha$ must be intrinsic to the 
secondary, probably the typical variations of active regions and starspots
as observed in single active stars.
 


\subsubsection{He I and Ca II Emission}

He I emission is a useful tool for the determination of the physical 
conditions in which it is
produced. The He I line at 5876\AA\ is a triplet while that at 6678 \AA\ is a singlet.
Three mechanisms populate these states: (1) recombination to an excited state after
photo-ionization by a $\lambda$ $<$ 504 \AA\ photons; (2) collisional excitation from the He I
ground state; (3) excitation of the singlets only by resonance scattering.
This latter option is easily eliminated here as we detect 
both singlets and triplets in emission.
Following the discussion related to the active M star AD Leo (Giampapa et al. 1978; Schneeberger
et al. 1979) it is shown that the I(5876\AA)/I(6678\AA) ratio can be as high as
45 in quiescent prominences in the Sun but is near 3.3 in active prominences. 
This latter value is near the ratio of the triplet to singlet
statistical weights (3.0). 

The I(5876\AA)/I(6678\AA) ratio of 45 has been explained 
(Heasley et al. 1974) as a
natural consequence of a typical quiet photosphere in a solar-like star. At cool temperatures
($<$ 8000K) the He I resonance line photons, which populate the singlet levels, can not
penetrate as far as the $\lambda$ $<$ 504 \AA\ photons which populate the triplet levels.
As the temperature increases, i.e., moving into an active stellar atmosphere, the
I(5876\AA)/I(6678\AA) ratio gets quickly reduced, almost independent of density, 
as collisional excitation from the ground state and collisional coupling of the triplet-singlet
levels bring the two populations into closer accord.
During low states, polars are not strong X-ray sources and even in high states, the X-ray
emission is primarily produced at the white dwarf magnetic pole 
accretion region(s). We can conclude that the excitation
of the secondary He I lines, during a low state, 
is not likely to be due to a strong $<$504\AA~continuum as needed for 
option (1) above. 
Further evidence of the lack of X-ray induced lines is provided by
the absence of He II or other high excitation emission lines in the 
optical spectra. 
Using the EF Eri spectra presented in Fig. 7, we can 
measure the I(5876\AA)/I(6678\AA) line
ratio\footnote{We can easily separate out the weaker Na emission sitting near
the He 5876\AA~line.}. 
We find I(5876\AA)/I(6678\AA) = 3.7, 
a value expected for He I emission by collisional
excitation in chromospheres hotter than 8000K. 
Athay (1965) provides rough guidelines for this
sort of line emitting region yielding column densities of $\sim$10$^{18}$ cm$^{-2}$ 
at temperatures of 20,000K or hotter.  

The Ca II IR triplet, in emission in EF Eri, has peak fluxes in the ratio 1:2.6:2
for the 8498, 8592, and 8662\AA\ lines, respectively. This is comparable
to the flux ratio observed in a wide variety of objects from CVs
(Persson 1988) to pre-main sequence stars (Hamann \& Persson 1992) to active
chromospheres in dMe stars (Pettersen 1989). The lines in the CVs are
fairly broad, exhibit disk
profiles, and form in the accretion disks; any weaker, narrower secondary
emission being hidden. The dMe stars
show narrow emission reversals in their Ca II photospheric absorption lines. The
pre-main sequence stars show both broad and narrow components. The broad
component seems to arise in a turbulent region near the base of the wind, while
the narrow component is at rest with respect to the star, and may form in the
chromosphere. In EF Eri the triplet lines are broadened by about 150 km/s,
which is somewhat larger than, but comparable to, the expected 
$\sim$100 km/s $v sin i$ of
the secondary. We suspect an origin in the chromosphere of the secondary.

Given the above discussion, we conclude the most likely source of the line emission
detected from the secondary in EF Eri is due to stellar activity.
We have observed that in EF Eri, compared with AM Her and VV Pup,
there was no secondary emission lines during the first 5-7 or so years of 
the low state but the emission began during the extended low state 
and has been present for
the past 1.5 years. This discovery argues for the fact that we have observed
the onset of a stellar activity cycle in the sub-stellar companion star in EF Eri.

\subsection{The Component Masses}

The white dwarf in EF Eri has been observed during the current low state by a few observers
(e.g., Beuermann et al. 2000, Howell 2004) who found it to be cool, 
near 9500K, and each made some
attempt to determine its mass. Beuermann et al. (2000) provided the most rigorous determination
for the mass of the white dwarf and concluded that the most likely 
value is near 0.6M$_{\odot}$. This value, equal to the mean mass for white dwarfs found in CVs,
has been generally used by the community for EF Eri. Harrison et al. (2002) fit optical and IR
photometry of EF Eri in the low state and found that an 0.6$M_{\odot}$ white dwarf worked quite
well in the overall spectral energy distribution for EF Eri and led to 
an estimated distance of 90 pc, a value in agreement with the determination by Beuermann et al. 

A recent paper by Wickramasinghe and Ferrario (2005) as well as former works by Liebert et al.
(2003) and Liebert (1988) has shown that the masses for isolated magnetic white dwarfs are, on
average, higher than non-magnetic white dwarfs. The mean masses of the two groups are
0.93M$_{\odot}$ and 0.57M$_{\odot}$ respectively. However, in cataclysmic variables the mean white
dwarf mass seems to be near 0.6M$_{\odot}$ regardless of magnetic or non-magnetic Warner (1995). 
Shylaja (2004) presents a compilation of derived masses for the white dwarfs in magnetic 
CVs and finds a trend toward higher mass (avg. near 0.85 M$_{\odot}$) but lower mass 
magnetic white dwarfs (0.4-0.6 $M_{\odot}$) constitute about 1/4 of the sample.
For now, we will adopt the value of 0.6M$_{\odot}$ for the mass of the 
magnetic white dwarf in EF Eri but examine the consequences of it having a higher 
mass later on.

Using the relationship between the size of the secondary 
star Roche lobe and the binary mass ratio
q(=M$_2$/M$_1$) and assuming the secondary object, regardless of its true nature, is centrally
condensed, we have
$$R_2/a = 0.462 (q/(1+q)) $$
(Paczynski 1971; Huber et al. 1998). The secondary Roche lobe size (R$_2$) for EF Eri is
$\sim$0.1R$_{sun}$$\sim$R$_{Jupiter}$ or 7$\times$10$^7$ meters 
(Howell et al. 2001) and the mean separation of
the two stars, $a$, is 3.5$\times$10$^8$ meters 
(Howell 2004)\footnote{Taking R$_2$ as equal 
to the Roche Lobe radius
implies that the secondary star fills the Roche lobe. During a low state this may or may not be 
a valid assumption (see Howell et al. 2000; Howell 2005).}. Substituting these values in the above
equation yields a mass
ratio of $q=0.092$ for EF Eri. Taking the white dwarf mass to be 0.6M$_{\odot}$, we find
M$_2$=0.055M$_{\odot}$. If the cutoff for core energy generation via hydrogen burning is taken
as 0.06M$_{\odot}$, then the secondary in EF Eri is sub-stellar and the binary may be a
post-period minimum system
(Howell et al. 2001). However, we should keep in mind the
use of a ``best guess'' white dwarf
mass and that our determined value for M$_2$, within errors, 
is essentially at the stellar/sub-stellar mass boundary. 

Taking the above masses at face value, our measured K$_2$ 
would predict a K$_1$ amplitude for the white dwarf of 25
km/sec, a value that will be extremely difficult to measure for the V=18.5 white dwarf in EF Eri,
especially given that its Balmer absorption lines are Zeeman split and 
their relative location and shape will probably 
change throughout the orbit as the effective observed magnetic field strength changes.
Of course, any precise K$_1$ measurement attempt could only be performed 
during a low state.

Beuermann et al. (2000) limited the secondary star in EF Eri to be 
later than M9 based on its
non-detection in the red optical spectral region and Howell and Ciardi (2001) used a moderately
good S/N K band spectrum to set a secondary mass limit of $\le$0.05M$_{\odot}$.
Using their spectral energy distribution, Harrison et al. (2004) concluded that the secondary
in EF Eri is near L5, a result that remained in agreement with the secondary's continued
non-detection in K band spectra (Harrison et al. 2005). Finally, Howell et al. (2006)
have used Spitzer Space Telescope observations of EF Eri in the 3-8 micron region to 
further set a limit on the secondary finding that it is consistent with 
an object approximating a very late L or T type brown dwarf.
All of these limits and mass estimates for the secondary 
in EF Eri seem to be in approximate
agreement, generally independent of an assumed
white dwarf mass. They all seem to indicate an M$_2$ value near 0.05-0.055M$_{\odot}$.
This mass for M$_2$ is also in agreement with the 
theoretical expectations for the secondary star in a CV such as EF Eri based on
its orbital period and the assumption of the standard 
CV evolution model (Kolb \& Baraffe, 1999; Howell et al. 2001; Politano 2004).

If the white dwarf in EF Eri has a higher mass, say as high as 
the average value for single
magnetic white dwarfs, 0.95 M$_{\odot}$, then M$_2$ would have a mass near
0.087M$_{\odot}$ assuming $q$ remains the same. While we cannot completely rule out this value for the mass of M$_2$,
such a star would not fit inside the secondary Roche Lobe if it were a normal main
sequence object. Therefore, at this mass, we'd expect the secondary to be a dense He
core with a thin hydrogen envelope, an object far less likely to show us what
appears to be the typical signatures of an active chromosphere.

\section{Discussion}


\subsection{Low States}

Today, we know that most polars, including EF Eri, AM Her, and
VV Pup have extended low states lasting years and that, in general, most polars show 
long term low states and/or
spend much of their time in low states (Gerke et al. 2006 and refs therein).
It is believed that all CVs (polars, dwarf novae, etc.) should have low states but 
that we only notice them in polars due to their lack of an accretion disk.
In CVs with disks, if the mass
transfer stops, the optical light may not show a significant dimming, 
as the optical light is dominated by or has a large contribution from
the accretion disk. The stoppage of mass transfer in a non-magnetic CV 
may not be easily noted and if mass
transfer restarts within a short time period ($\sim$2-4 weeks), 
the entire event might escape detection.
King and Cannizzo (1998) provide a theoretical framework for this idea and explore possible
observational ramifications that could result. They conclude that current observations do not
not agree with their predictions of how a disk system would react to a stoppage of mass
transfer. Given their work and the fact that polars often show extended low states
it is hard to reconcile the idea that CVs with accretion disks (i.e.,
the non-magnetic white dwarf CVs) have low states similar to
polars. The VY Scl stars may be the exception, as we discuss below.

It has been argued that the low states are a result of 
stellar activity cycles on the secondary
star, in particular starspots migrating to the L1 region
and stopping mass transfer for some period of time. 
The idea that ``solar cycles" and starspots cause low states has been around for decades. 
Van Buren and Young (1985) suggested that changes in radius of the 
magnetically-active secondary in RS CVn systems drove the observed orbital
period variations. The basic idea is that, during a stellar activity cycle, 
the magnetic pressure due to the enhanced subsurface magnetic fields 
displace gas, resulting in a fractionally larger stellar radius. The
change in the moment of inertia drives the system out of synchronous rotation,
and tidal torques then quickly bring the system back into synchrony. In the
case of the RS~CVn secondaries, the fractional change in the orbital periods
($\sim$10$^{-6}$) requires a similar fractional change in the stellar radius.
Applegate \& Patterson (1987) applied this type of model to the cataclysmic
variables, and predicted that there should be periodic $O-C$ variations on the 
magnetic activity period of the secondary. These magnetic activity periods
are observed to be of order a decade long, within a factor of 2 of the
length of the Solar cycle. 

In the case of the polars, it may be this fractional change in the stellar
radius that comes into play. Since the secondary is filling its Roche lobe
when in the high state, a small reduction in the stellar radius as the
magnetic activity wanes may lead to a cessation of the accretion. The
implication is that as the magnetic activity picks up, the star should
then expand and eventually accretion should resume. This seems consistent with
what we observed in the EF Eri secondary: the magnetic activity was at a
minimum when the accretion ceased, and then increased prior to the start of
the current high state.

The few direct observational results we do have that reveal stellar activity are
all from polars in low states.
We have seen that EF Eri entered a low state and its secondary was in-active, 
it turned on, and
EF Eri stayed in a low state for another 1.5 years. 
On the other hand, AM Her and VV Pup
with normal active M star secondaries, went
into low states (recently, twice each) and from start to end during the low state 
the secondary
was active, at a constant level, throughout. 
Single active M stars display spectroscopic activity indicators almost all the time but
when they go into flaring or super-active episodes, their chromospheres 
can expand locally and eject material.
This idea of active chromospheric expansion driving
L1 mass loss was proposed by Howell et al. (2000) as the possible cause of
high/low state. The model removed the need for a starspot at L1 but still 
kept the idea that stellar activity/flaring somehow triggers state changes.
If some magnetic cycle connection is invoked to drive mass flows, then
it may be that even in a high state, the photosphere need not fill 
its Roche lobe
The evidence for starspots at L1
and/or stellar activity {\it directly} starting and stopping high/low states seems not to exist.

Assuming, for the moment, that non-magnetic CVs do not undergo low states, can we
explain such a phenomena? 
Polar secondaries have been shown to be normal main sequence stars (Harrison et al. 2005)
while dwarf nova secondary stars, to date at least for systems above the period gap,
have been shown to have spectra that reveal
odd abundance patterns and are consistent with CNO processed material.
Evidence has been presented that argues for a divergence in the
evolutionary history of polars and non-magnetic CVs
(see Schmidt et al. 2005; Harrison et al. 2005).
If the secondaries in non-magnetic CVs are in fact remnant He cores
from a past time when they were more massive, as has been suggested
by Beuermann et al. (1998), Howell (2005), and Harrison et al. (2005), possibly
evolving from Algols, then maybe normal stellar activity is not possible.
If high/low states are somehow related to 
activity cycles on the mass donor, this idea may provide a natural explanation as to why 
dwarf novae and other non-magnetic CVs do not have polar-like low states.

Having said this, we must consider the VY Scl type of CV. These systems are believed to have non-magnetic white dwarfs and accretion disks, yet they show low state behavior
that may be similar to that seen in polars. 
VY Scl stars have been used as ``proof" of starspots and stellar activity but to date,
this is more speculation than anything else.
This sub-class of CV lies in the 3-4 hour period range, directly above the
period gap. Howell et al. (2001) suggests these stars behave as they do because their
secondaries are far from thermal equilibrium and mass transfer can be highly modulated by
thermal timescale changes caused by mass loss. If this is true or not, is yet to be seen,
but observations of the secondary in VY Scl stars are likely to
offer important clues as to the cause of high/low states. 

\subsection {Related Binaries}

What does the secondary star do in terms of its activity cycles
before, during, and after high/low state transitions? 
We have seen that stellar activity (as indicated by the usual optical emission
lines) may be a constant feature in the normal M star
secondaries of polars (i.e., VV Pup and AM Her) and that it 
can ``turn on" during a low state (or at other times?)
in at least one sub-stellar mass object, the secondary in EF Eri. 
Let us examine a few other binary systems in which magnetic fields 
and stellar activity play important roles.
These objects may provide valuable clues toward our understanding of polars.

\subsubsection {RS~CVn Binaries}

Low mass stars in close binaries exhibit enhanced activity, which is 
generally attributed to the amplification of the magnetic dynamo by
tidally-enforced rapid rotation. The activity is generally
saturated, with activity-related emission at the level of
10$^{-2}$ to 10$^{-3}$ of the bolometric flux. The secondaries
of polars have similarly-rapid rotation and, if magnetic dynamo 
processes are present they may also be expected to exhibit 
activity at the saturated levels. However, the analogy is not exact,
as the polar secondaries may have lost a significant fraction of
their convective zones, with an effect on their ability to maintain
or amplify a magnetic dynamo.

Stellar magnetic field emergence is known to be concentrated at certain active
longitudes, both in the Sun and more active stars. 
The active RS~CVn binaries are characterized
by a ``photometric wave", interpreted as a spotted (and magnetically
active) hemisphere that 
migrates around one of the stars with respect to the binary phase
with a period of order a decade.
The stars are tidally locked in such close binary systems; the migration
is interpreted to represent differential rotation of the star, as the starspots
migrate equator-ward as they do during a solar cycle.  
While there is evidence for such a photometric wave with a 6-9 month
period in the pre-cataclysmic variable V471 Tau
system (\.{I}banogl\u{u} et al.\ 1994), there is also evidence that the
chromospheric H$\alpha$ is enhanced
near the substellar longitudes (Rottler et al.\ 2002). Because the strength
of the emission decreased with time, Rottler et al.\ dismissed irradiation
as the cause, and suggested that there is a permanent active longitude at the
substellar point, caused by tidal distortion of the convective dynamo.
It is not clear whether this is in conflict with the observed wave migration
(see \S4.2.2).

There is spectroscopic evidence for enhanced flaring activity at the 
substellar
longitudes in the RS~CVn system UX~Arietis (Simon, Linsky, \& Schiffer 1980),
and photometric evidence for enhanced flaring activity at the substellar
longitudes in the close binary dMe system YY~Gem (Doyle \& Mathioudakis 1990).
This is presumably attributable to recombination between the extended magnetic
loops of the two stars, as in the models by Uchida and Sakurai (1985) 
(see Figure 11). There is no
evidence addressing the long term variability of the flaring rates. If the
recombination involves 
large-scale loops with sizes of order the binary separation, then flaring rates
may remain more-or-less constant, but the rates may be
enhanced when the active longitudes coincide with the substellar point.

Eclipse-mapping observations of RS CVn systems, of Algol, and of YY~Gem have
been brought to bear on the question of the presence of magnetic
connections between stars, as might be expected if the substellar magnetic
activity enhancement is indeed permanent.
Pre\'s, Siarkowski, \&  Sylwester (1995) and Siarkowski et al.\ (1996)
modeled X-ray eclipse observations of the RS CVn systems TY~Pyx and AR~Lac,
respectively. They concluded that in both cases 
a significant fraction of the emission arises between the stars, in
magnetic loops connecting the stars.
However, their unconstrained maximum entropy
modeling solutions are not unique, and other equally good solutions
without inter-stellar emission exist (see the review by G\"udel [2004]).

\subsubsection {V471 Tau}

For V471 Tau, often used as the prototype pre-CV, the hot white dwarf
(T=35,000K) was originally thought to irradiate the K secondary star.
The evidence for this was H$\alpha$ emission from the secondary star 
which was observed to be
concentrated toward the white dwarf with an equivalent width that peaked
near orbital phase 0.5 and the emission line disappeared completely 
when looking at the
back side  of the secondary. However, multi-year observations of V471 Tau by
Rottler et al. (1998, 2002) showed that in 1987, the H$\alpha$ emission
was consistent with an irradiation interpretation while in 1990 the emission
was much weaker and heavily concentrated toward the white dwarf and in 1992 the
emission was completely absent. The (non-magnetic) white dwarf in V471 Tau
was observed to be constant throughout this entire 5 year period showing
that the secondary star emission was not due to irradiation by the hot 
white dwarf. Rottler et al. believe their observations are consistent with a
``solar cycle"-like change that occurred in the K star.

Given their result with V471 Tau, Rottler et al. (2002) (re)examined a number of
similar WD + RD systems all of which were supposed to show irradiation induced
emission from the non-interacting secondary star. Of the ten hot white dwarf
(T=30,000 to 60,000K) plus red dwarf stars, only one (NN Ser, a WD+RD pair with T$_{WD}$=55,000K) 
is probably a true case of irradiation. The others are shown to have 
not enough UV flux to cause irradiation and the
variable H$\alpha$ emission was
consistent with activity induced emission lines.
Using a model based on the number of $<$912\AA\ photons available for
irradiation of the secondary, they show that an incident UV flux at the surface
of the secondary star of $\la$1$\times$10$^{10}$ ergs sec$^{-1}$ cm$^{-2}$
ster$^{-1}$ is not sufficient to produce irradiation induced emission lines.
This value exceeds that present in EF Eri from its cooler but closer 
white dwarf
(F$_{UV}$ $\sim$1$\times$10$^{9}$ ergs sec$^{-1}$ cm$^{-2}$ ster$^{-1}$) 
and, in fact, it exceeds nearly every CV when not in outburst.

\subsubsection {SDSS J121209.31+013627.7}

It is interesting to note here the discovery of an object that 
is very similar to
EF Eri in the low state, SDSS J121209.31+013627.7 (Schmidt et al. 2005). These authors
discuss this 90 minute, cool (T$\sim$10,000K), orbitally 
synchronized magnetic (7-13 MG) white dwarf binary
and present optical spectroscopy closely approximating those of EF Eri in a low
state. They
conclude, however, that SDSS J1212's H$\alpha$ emission is most likely 
due to irradiation of the probable brown dwarf secondary based on a single
epoch set of phase-resolved optical spectroscopy. 
The H$\alpha$ emission
completely disappears when viewing the back end of the secondary star,
a similar result to that observed once in V471 Tau.
We therefore can ask, given the nearly similar nature of J1212 and EF Eri,
if the secondary in J1212 is irradiated, 
why isn't the secondary in EF Eri? 
With stellar activity seemingly being concentrated on the 
secondary at the sub-stellar point of the white dwarf, 
we will need to be careful in our
observational interpretation of any detected emission from the secondary. Radiatively heated
atmospheres (irradiation) and active stellar chromospheres (starspots etc.)
may be hard to distinguish. 
If a star is irradiated, the apparent spectral type (photospheric temperature) 
should vary
as the star rotates. Active chromospheres do fill in lines, which mimics
an earlier spectral type in the K stars, but the veiling is
wavelength-dependent. In M stars, the veiling might give a spectral type
earlier than T$_{eff}$, since molecular band strengths are increasing with
decreasing T$_{eff}$.
Realistic, 3-D, magneto-hydrodynamic models of the secondary star and its 
magnetic, tidal, and radiative interaction with the
primary are needed to fully understand the observations.
J1212 and EF Eri are good starting points to use as proxies 
for our understanding 
of irradiation, stellar activity on brown dwarf-like secondary stars, and
exoplanet physics. With no mass transfer currently underway in J1212, this
system is an ideal candidate for multi-epoch observations to monitor and detail
the nature of the secondary star emission lines.

\subsubsection{Extra-Solar Planets}

The analogy between polars and the magnetically-active binary systems is
imperfect.
In none of these cases does the active star fill its Roche lobe, in none of
these cases does the strength of the photospheric magnetic field exceed a
few kG, and in none of these cases is the secondary star bathed in 
a strong external magnetic field. The 
tidal forces are also much stronger in EF Eri than even in V471 Tau, where the
period is 8 times longer and the mass of the cool star is over an
order of magnitude larger. A better analogy may be between a star and a planet.

Perhaps the answer to starting and stopping mass transfer lies in the area of 
magnetic (re)connection between the white dwarf and the secondary star.
It was observed (Shkolnik et al. 2003) that for  ``hot Jupiter" type extra-solar 
planets, the host star Ca II H\&K emission lines are sometimes modulated on the orbital period of the
planet. A model proposed by Ip et al. (2004) explains this phenomenon as an 
interaction of the exoplanet magnetosphere with that of the parent star, 
in which a magnetic flux loop reconnects
using the nearby planet as a conductor. In a polar, which has a much stronger 
field, it seems obvious that magnetic reconnection and closed field loops
would have to pass through the secondary. This idea may also explain why the 
onset of stellar activity alone is not the direct cause of high/low states and 
why polars in low states seem to have a residual amount of mass transfer, 
probably magnetically connected secondary star wind accretion.

A possible observation of magnetic reconnection and closed field loops in 
the region near L1 but between the two stars (as illustrated in Fig. 11) 
was noted by Kafka et al. (2005, 2006)
in their low state observations of stellar activity in AM
Her. The H$\alpha$ emission line profile is triple peaked and leads to a model in which 
the regions on the secondary star where stellar activity occurs seem to 
be preferentially on the side facing the white dwarf and 
contain loop type structures surrounding the secondary. 
A similar white dwarf facing activity concentration
and a similar multi-peaked H$\alpha$ line has been observed in V471 Tau 
(Young et al. 1991). 
The H$\alpha$ emission line satellites have been stable for over 2 years in
AM Her and VV Pup showed a similar H$\alpha$ emission line structure 
during one of its recent low states (Mason et al. 2006). 

\subsection {Additional Observational Study}

Stellar activity, 
as evidenced by emission lines such as H and Ca, is usually
an optical bandpass specific proxy. Even very active stars, 
(see Fig. 12),
show little obvious evidence for chromospheric activity in their 
near-IR and IR spectra.  The reason that activity emission is so weak in the near-IR and IR bands 
is both a contrast effect and one of line formation. The activity 
induced emission lines in an active binary are often
``filled in" by the emission from the bright photospheric continuum.
Additionally, the typical spectral lines present in the IR region 
that one might expect to be in emission and associated
with stellar activity (e.g., $J$- to $K$-band H I Paschen series and Ca lines) 
form too low in the stellar atmosphere to be greatly
modulated or affected by an active chromosphere.
In CVs, the contrast effect just mentioned can be provided by the high state
accretion flux or accretion disk light, hiding not only the secondary but any
possible activity-induced lines. Additionally, CV emission lines are 
generally very broad due to the high velocities in the disk or stream,
again able to hide weaker, narrow lines caused by activity.
Polars in low states offer the lack of additional (accretion) flux contribution
allowing secondary star activity indicators to be observed in the optical.
However, formation of the higher energy IR lines deeper in the stellar 
atmosphere still renders the 1-3 micron region a poor choice in the search for
chromospheric activity.
Indeed, near-IR spectroscopy of EF Eri
obtained after the optical emission lines appeared (Johnson et al. 2005), 
show no sign of hydrogen or any other emission lines. 
ST LMi as well, was observed at 2.2
microns during a low state and no emission lines were seen, although 
Balmer emission was present (Howell et al. 2000).
Thus, searches for stellar activity in CV secondaries are 
probably limited to low state observations in
the optical, and as such, are mainly restricted to polars.

Low state observations, such as those presented herein, are generally 
difficult to gather as they
require optical spectroscopy, often as target of opportunity observations,
with fairly large telescopes, as low state polars tend to be faint.
Phase resolved spectroscopy is required as the origin and cause of the 
spectral emission (or absorption, see Mason et al. 2006) features observed must
be firmly determined. A single spectrum or even a single epoch of 
observations can easily confuse activity with irradiation.
Polars are probably the easiest targets for this
purpose as they show their 
high/low states directly while the VY Scl stars might be considered
prime targets to go after as we know little about their low states or their
secondary stars.

We have provided new direct spectroscopic evidence into the mix 
of understanding high/low states of polars and if and how these mass transfer 
changes are related to stellar activity. The
simple idea of the turn on of stellar activity {\it directly} starting 
and stopping mass transfer 
(i.e., high and low states) seems ruled out as activity 
turned on in EF Eri, yet it  remained low for another
1.5 years, while stellar activity seems ever present in the secondary during 
low states of the polars
AM Her and VV Pup. Just as we finish this paper, we can report that EF Eri
has reentered a high state, reaching V=15.6 on 
2006 March 04 (Stubbings, priv.  comm.)\footnote{ 
Note that there seems to be nothing special about the location of 
the start of the high state (JD 2453798.) in Fig. 10.}.
Data obtained up to the time of this rebrightening (see Figs. 8, 9, 11)
reveal no obvious change in the radial velocity, emission distribution on 
the secondary, 
or the EW of H$\alpha$ directly before the high state started.
Our monitoring programs show EF Eri to be currently providing all its 
usual high state properties observed in the past, even 
after its nine year hiatus.

\acknowledgments
We want to make a special mention here of the undaunted observational
effort of Rod Stubbings who, for over nine years, observed EF Eri
and provided us with nothing, nothing to report that is, on EF Eri being visible
until a few weeks ago when his email subject line ``EF Eri brightened !!!"
got the attention of astronomers the world over. Thanks Rod!
We want to acknowledge 
Sandy Leggett \& Tom Geballe for obtaining the K band images of
EF Eri and UKIRT for its continued service observing program. 
Margeret Hanson kindly obtained the spectra of RS CVn for us.
Frank Hill and Mark Giampapa have generously contributed their time to 
converse on and provide a great resource for information
related to stellar activity on solar-like stars. We thank the referee 
for providing a number of useful comments.
We are grateful for the support of Dean of Arts and Sciences J.~Staros,
Provost R.~McGrath, and Vice President for Research G.~Habich, all of
Stony Brook University, for providing partial support that enabled
Stony Brook's participation
in the SMARTS consortium. We thank the SMARTS service observers, J. Espinoza,
D. Gonzalez, and A. Pasten, for taking the data,
and for their dedication to the SMARTS effort.
We thank C. Bailyn, the driving force behind the SMARTS consortium, for his
leadership. This research was funded in part by NSF grant AST-0307454
to Stony Brook University.
Some of the observations used herein were obtained at the W. M. Keck 
Observatory, 
which is operated as a scientific partnership among the California 
Institute of Technology, the University of California and the National 
Aeronautics and Space Administration, made possible by the generous 
financial support of the W. M. Keck Foundation.
The authors wish to recognize and acknowledge the very significant 
cultural role and reverence that the summit of Mauna Kea has always had 
within the indigenous Hawaiian community.  We are most fortunate to have 
the opportunity to conduct observations from this mountain.
Part of this work was performed (M.E.H \& R.H.B) under the auspices of the U.S.
Department of Energy, National Nuclear Security Administration by the
University of California, Lawrence Livermore National Laboratory
under contract No. W-7405-Eng-48.
We wish to that the IAU for permission to reproduce Figure 10.
Finally, SBH wishes to thank ESO headquarters in Chile for their support, 
and my host, Dr. E. Mason, for her coffee and cooking which provided 
a wonderful working environment, allowing the completion of this paper.

\clearpage

\begin{deluxetable}{cccc}
\tablenum{1}
\tablewidth{6.5in}
\tablecaption{SMARTS Spectroscopy Observing Log}
\tablehead{
 \colhead{HJD of Mid-exposure}
 & \colhead{Exp.Time (sec)}
 & \colhead{Spec. Phase\tablenotemark{a}}
 & \colhead{H$\alpha$ Center (\AA)}
}
\startdata
\multicolumn{4}{c}{2005 Nov 03 UT} \\
 2453677.59576  & 600 & 0.5281550 & 6564.534 \\
 2453677.60287  & 600 & 0.6545194 & 6559.301 \\
 2453677.61000  & 600 & 0.7812392 & 6559.816 \\
 2453677.61709  & 600 & 0.9072481 & 6565.360 \\
 2453677.62420  & 600 & 0.0336125 & 6562.128 \\
 2453677.63843  & 600 & 0.2865190 & 6569.461 \\
 2453677.64047  & 600 & 0.3227754 & 6567.428 \\
 2453677.65265  & 600 & 0.5392477 & 6563.449 \\
 2453677.65976  & 600 & 0.6656121 & 6559.426 \\
 2453677.66685  & 600 & 0.7916210 & 6557.741 \\
 2453677.67396  & 600 & 0.9179854 & 6560.756 \\
\multicolumn{4}{c}{2005 Dec 12 UT} \\
 2453716.57245  & 600 & 0.2517711 & 6568.204 \\
 2453716.57955  & 600 & 0.3779577 & 6568.298 \\
 2453716.58668  & 600 & 0.5406775 & 6564.490 \\
 2453716.60089  & 600 & 0.7572286 & 6558.563 \\
 2453716.60799  & 600 & 0.8834152 & 6566.133 \\
 2453716.61512  & 600 & 0.0101350 & 6566.893 \\
 2453716.62222  & 600 & 0.1363217 & 6567.989 \\
 2453716.62933  & 600 & 0.2626860 & 6569.877 \\
 2453716.63644  & 600 & 0.3890504 & 6571.038 \\
 2453716.64354  & 600 & 0.5152371 & 6565.521 \\
 2453716.65064  & 600 & 0.6414237 & 6559.827 \\
 2453716.65774  & 600 & 0.7676103 & 6558.431 \\
 2453716.66487  & 600 & 0.8940302 & 6562.807 \\
 2453716.67201  & 600 & 0.0212277 & 6570.044 \\
\multicolumn{4}{c}{2005 Dec 24 UT} \\
 2453728.60122  & 600 & 0.0362933 & 6559.763 \\
 2453728.60833  & 600 & 0.1626577 & 6573.287 \\
 2453728.61544  & 600 & 0.2890221 & 6569.500 \\
 2453728.62254  & 600 & 0.4152087 & 6567.619 \\
 2453728.62965  & 600 & 0.5415731 & 6563.255 \\
 2453728.63677  & 600 & 0.6681152 & 6557.919 \\
 2453728.64387  & 600 & 0.7943018 & 6558.931 \\
 2453728.65099  & 600 & 0.9208439 & 6563.869 \\
\multicolumn{4}{c}{2006 Feb 05 UT} \\
2453771.55291 &  600 & 0.4683228 & 6570.381 \\
2453771.55975 &  600 & 0.5899048 & 6564.660 \\
2453771.56685 &  600 & 0.7160645 & 6560.325 \\
2453771.57396 &  600 & 0.8424683 & 6558.134 \\
2453771.58106 &  600 & 0.9686279 & 6565.417 \\
2453771.58817 &  600 & 0.0949707 & 6570.402 \\
2453771.59530 &  600 & 0.2217407 & 6573.849 \\
2453771.60241 &  600 & 0.3480835 & 6570.223 \\
2453771.60950 &  600 & 0.4740601 & 6567.156 \\
2453771.61661 &  600 & 0.6004639 & 6563.936 \\
2453771.62374 &  600 & 0.7271729 & 6559.603 \\
2453771.63084 &  600 & 0.8533325 & 6558.002 \\
\multicolumn{4}{c}{Other Measured Spectra} \\
 2453298.61870  & 600 & 0.1806642 & 6570.467 \\
 2453298.62590  & 600 & 0.3085938 & 6569.269 \\
 2453298.63500  & 900 & 0.5014648 & 6564.754 \\
 2453330.57980  & 600 & 0.2177734 & 6571.960 \\
 2453330.58620  & 600 & 0.3315430 & 6569.405 \\
 2453579.89020  & 500 & 0.1430664 & 6570.897 \\
 2453579.89620  & 500 & 0.2497559 & 6571.325 \\
 2453579.90220  & 500 & 0.3564453 & 6569.764 \\
 2453620.72200  & 1200 & 0.9091797 & 6557.291 \\
 2453620.73610  & 1200 & 0.1597900 & 6569.510 \\
 2453620.75020  & 1200 & 0.4104004 & 6568.200 \\
 2453621.75480  & 1200 & 0.2648926 & 6569.407 \\
 2453621.76890  & 1200 & 0.5155028 & 6566.310 \\
 2453621.78300  & 1200 & 0.7661133 & 6556.565 \\
 2453635.73010  & 1200 & 0.6446533 & 6562.242 \\
 2453689.63210  & 600 & 0.5705872 & 6563.307 \\
\hline
\enddata
\tablenotetext{a}{Determined using the new spectroscopic ephemeris given herein.
Phase 0.0 equals the time of primary (white dwarf) red-to-blue crossing which equals
the time of secondary inferior conjunction. Spectroscopic phase 0.0 =
Bailey et al. photometric phase 0.41.}
\end{deluxetable}


\clearpage

\figcaption[]{1a - EF Eri has remained stubbornly inactive during this 2.3 year interval
as evidenced by the presented SMARTS photometry.
  The cluster of points at JD 3677 in the V band represents a 90 minute,
  full orbit campaign as discussed in the text. 
These differential magnitude light curves are plotted as usual, that is 
smaller numbers being brighter.
The magnitude uncertainties are dominated by counting statistics
in EF Eri.
1b - The differential photometry from 22 August 2003 through 21 December 2005,
shown in Fig. 1a,
  folded on EF Eri's orbital period and new spectroscopic 
ephemeris as given in the text. The light curve half-amplitude
  is 0.09 and 0.08 mag in the B and V bands, respectively, with no
significant modulation in the I band. This long term light curve 
is in good agreement
with the single orbit light curve presented in Fig. 2 of Harrison et al. (2003), the latter
being phased on the Bailey ephemeris.
The differential light curves are plotted as usual, that is 
smaller numbers being brighter.
}

\figcaption[]{Our SMARTS spectrum of EF Eri obtained on 19 Oct 2004 UT.
The Balmer and other emission lines are obvious as well as the underlying
Zeeman split absorption lines due to the white dwarf. The emission lines appear
to be weaker than those observed in Jan 2006 (see Fig. 7), but the 
spectral resolution is 17\AA, so the lines may simply be weaker relative to the
continuum. The spectrum is flux calibrated and presented in arbitrary units.
}

\figcaption[]{
The radial velocity of the H$\alpha$ emission line and the 
simultaneous $V$-band photometry from SMARTS on 3 November 2005.
The radial velocities are from a Gaussian fit of the emission feature; 
the uncertainties represent a $\pm$1~\AA\ systematic uncertainty due
to velocity smearing and the skewing due to a non-flat underlying
continuum.  This shows the relation between the photometric
orbital modulation of the white dwarf and the orbital location of the
secondary.
}

\figcaption[]{Representative SMARTS spectra of EF Eri. The top three spectra are single
10-minute exposures while the bottom spectrum (thick line) is their sum. 
The vertical dotted line represents the rest velocity of H-alpha.
     Fluxes for each spectrum are on the same scale, but are offset by 
one unit from the previous spectrum.
Compare these
observations with the Keck spectrum of EF Eri (Fig. 9) and you'll see that the SMARTS data
allows us to measure the top of the H$\alpha$ profile.
}

\figcaption[]{A trailed spectrum of EF Eri for the observations 
obtained on 4 Feb 2006.
The slightly jumpy nature of the H$\alpha$ emission line is
mainly due to limited phase sampling of the ultra-short 81 minute binary orbit.
Changes in the line strength are due to small seeing variations but also 
have a real component
as discussed in the text (See Fig. 9). The data are not smoothed and 
the bright spot near phase 0.8 at $\sim$6700\AA\ is a cosmic ray.
}
\figcaption[]{A comparison of the low state blue spectra of AM Her and EF Eri.
These polars have white dwarfs with similar magnetic field strengths (near 13 MG)
but different white dwarf temperatures (20,000K for AM Her, 9500K for EF Eri).
Their orbital periods are different as well, 3 hours vs. 81 minutes.
Note the presence in both stars of Ca II H\&K emission as well as how similar the
Zeeman split white dwarf Balmer lines are.
These spectra were obtained in Oct. 2005 at the 3.5-m APO telescope and
had integration times of 20 minutes each, a value that is one-third of
EF Eri's orbital period (P$_{\rm orb}$ = 81 min). The ordinate axis is correct for AM Her, while EF Eri has been
multiplied by ten.
}
\figcaption[]{EF Eri as observed by Keck II in Jan 2006. The spectrum covers 
the majority of
the optical band pass at quite good S/N (full range shown at top left) 
and reveals the set of emission lines from the
secondary (see expanded views). 
The underlying white dwarf Balmer absorption lines are Zeeman split and reveal the
surface magnetic field strength. See text for details.
}
\figcaption[]{The H$\alpha$ radial velocity curve for EF Eri 
based on SMARTS spectroscopy. The solid line is our
best fit as described in the text. The top axis shows the photometric
phase as determined by Bailey et al. (1982) and the bottom axis show our new
spectroscopic orbital phase as described in the text. Each of the group one nights are shown
in a different color: Red = 2005 Nov 03, Green = 2005 Dec 12, Blue = 2005 Dec 24, 
Orange = 2006 Feb 05. The purple points show our remaining monitoring 
RV measurements (see Table 1). See the text for details.
}

\figcaption[]{The phase resolved equivalent width of 
the H$\alpha$ emission line 
for the four group one nights in Nov. and Dec. 2005 and Feb 2006. 
Each orbit is shown by a different color 
(and symbol; 2005 Nov 03 is a filled square, 2005 Dec 12 is a filled star,
2005 Dec 24 is a filled triangle, and 2006 Feb 05 is a filled circle) 
following the scheme used in Fig. 8. The data are phased on the
spectroscopic ephemeris presented in the text and each measurement has an
approximate uncertainty of $\pm$2\AA. The eq. width of the line is essentially
constant at all phases with a possibly weaker line during phases 0.9 to 0.1,
the back side of the secondary.
}

\figcaption[]{
  The long term variation of the H$\alpha$ equivalent width integrated over
the full observation (0.4 - 1.2 orbital cycles).
   The equivalent width is plotted (i.e., the line is in emission) and 
  the units of W$_{\lambda}$ are \AA .
}
\figcaption[]{Magneto-dynamical model of the magnetic field interactions between 
the two component stars in an RS CVn binary. RS CVn stars are highly
chromospheric active binaries
often containing a KIII (top star here) having a polar magnetic field near 
1000G and a main sequence F-G star having a polar field of $\sim$100G. The stars
orbit in a few to tens of days and have their rotation locked to the orbit.
Changes in the magnetic field structures, due to starspot migration via
differential rotation, causes flux tube connection, breaking,
and reconnection events. Direct magnetic pole to starspot field lines exist
and the majority of the chromospheric activity in the K star occurs on the 
companion facing hemisphere. One can imagine a similar but greatly enhanced
process might occur in polars with the secondary star activity concentrated
toward the white dwarf. Figure reproduced with permission from 
Y. Uchida and T. Sakurai, "Magnetodynamical Processes in
Interacting Magnetospheres of RS CVn Binaries", published in IAU Symposium
Series n. 107 "Unstable Current Systems and Plasma Instabilities in
Astrophysics", pp. 281-285, 1985, D. Reidel Publishing Co.
}

\figcaption[]{IRTF SPEX spectra of the highly active stars (a) RS CVn and 
(b) HD28867S. RS CVn, a short period active binary system, 
presents an example of a contrast issue as the giant
companion's photosphere fills in the activity 
induced emission in the near-IR (i.e., Ca II triplet) 
but we do observe a weak He 10830\AA~emission line.
In HD28867S, an M2 pre-main sequence star (Walter et al.\ 2003),
we again see a He 10830\AA~emission line, but with a P~Cygni profile. The
emission is at zero velocity; the absorption is likely due to accretion
in this young star. H~I Paschen~$\gamma$ is weakly in emission, with an
equivalent width of -0.4\AA. The Ca II IR triplet is in absorption.
While both stars show strong Ca II H\&K and
H$\alpha$ emission in the optical, their infrared spectra do not contain a similar
wealth of strong
emission lines generated by stellar magnetic activity at temperatures of
6000-20000K. This is due both to contrast against a bright photosphere
near the peak of the Planck function, and the depth of the IR line
formation in stellar atmospheres.
}

\newpage

\begin{figure}
\plotone{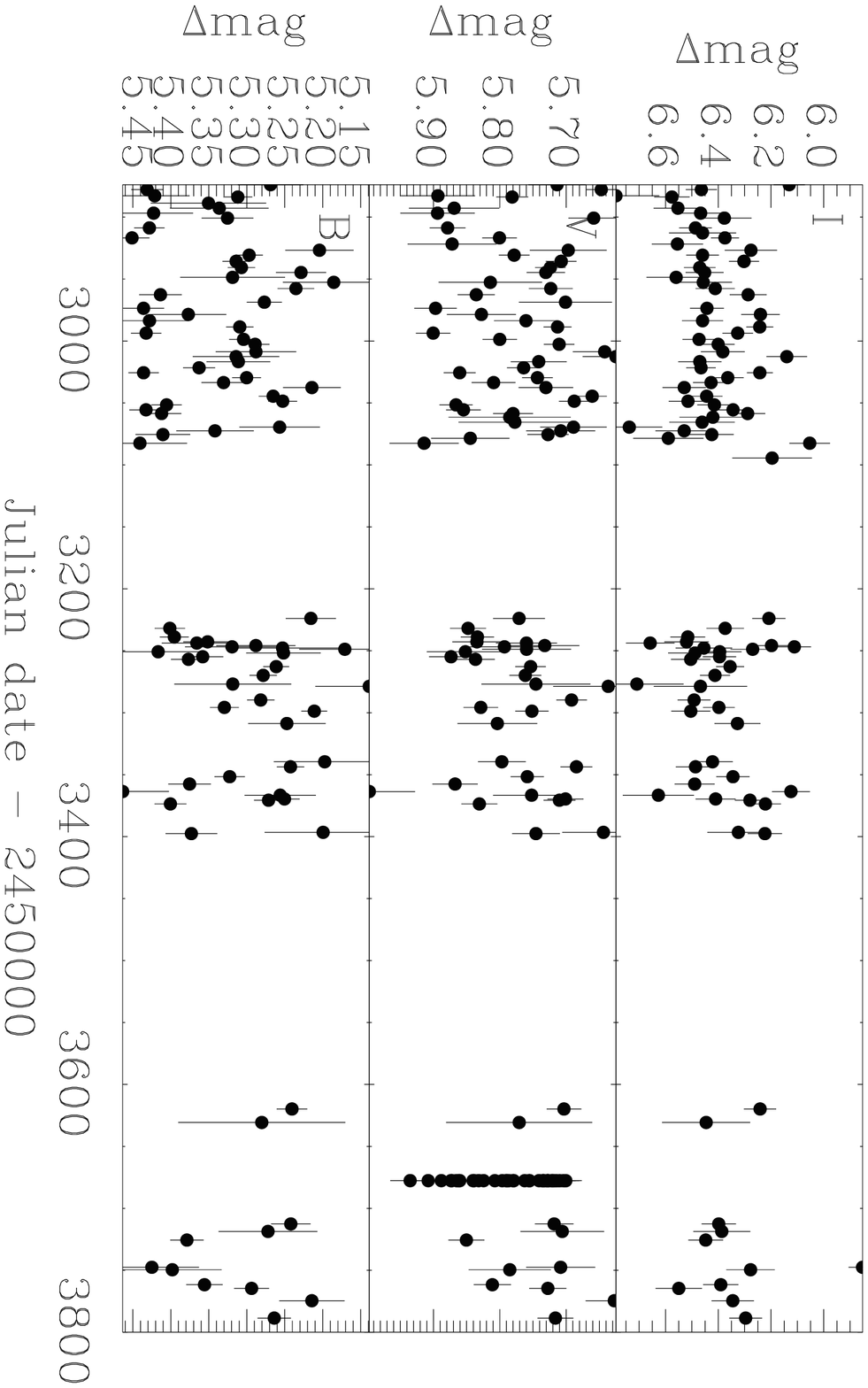}
\end{figure}

\begin{figure}
\plotone{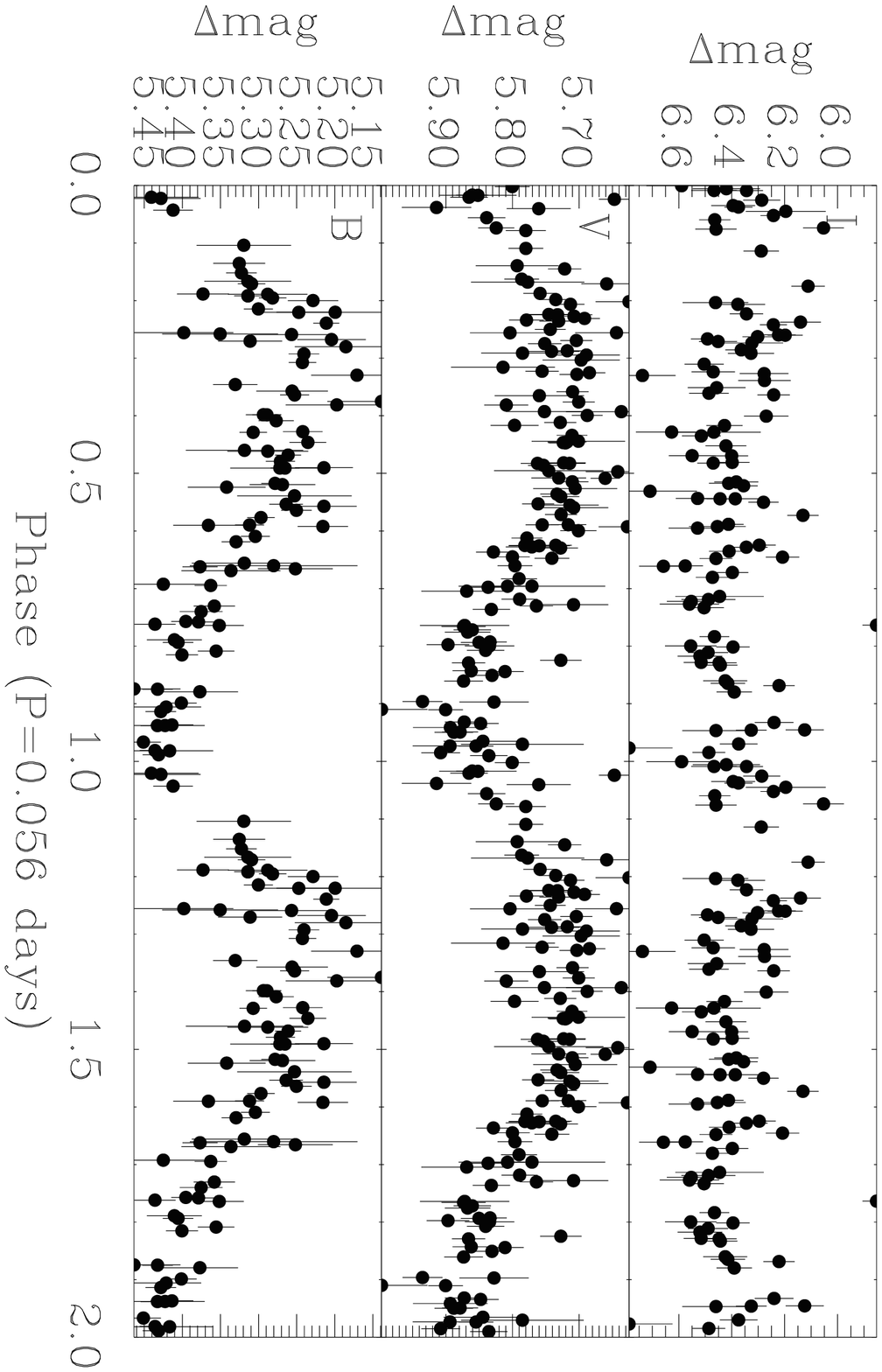}
\end{figure}

\begin{figure}
\plotone{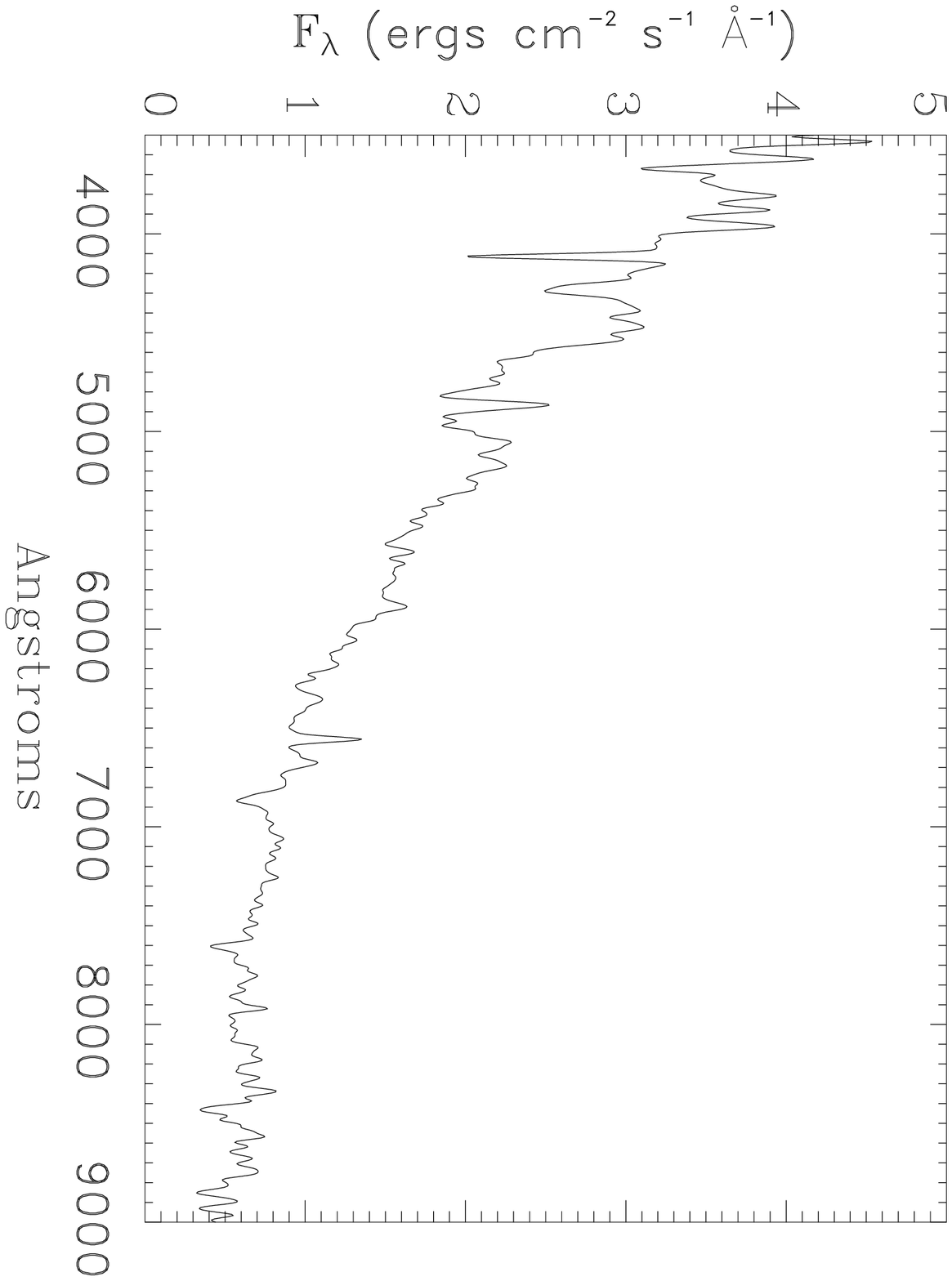}
\end{figure}

\begin{figure}
\plotone{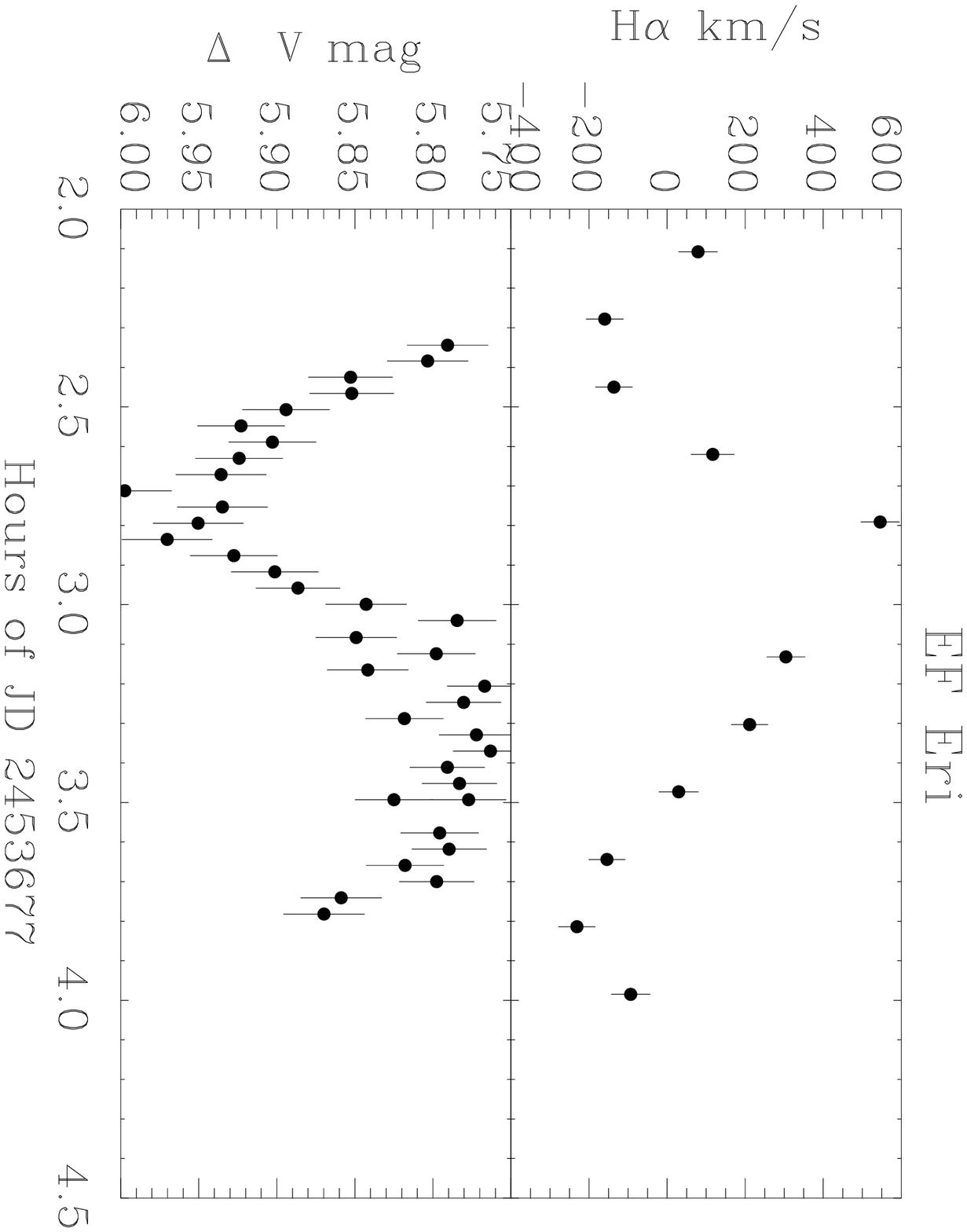}
\end{figure}

\begin{figure}
\plotone{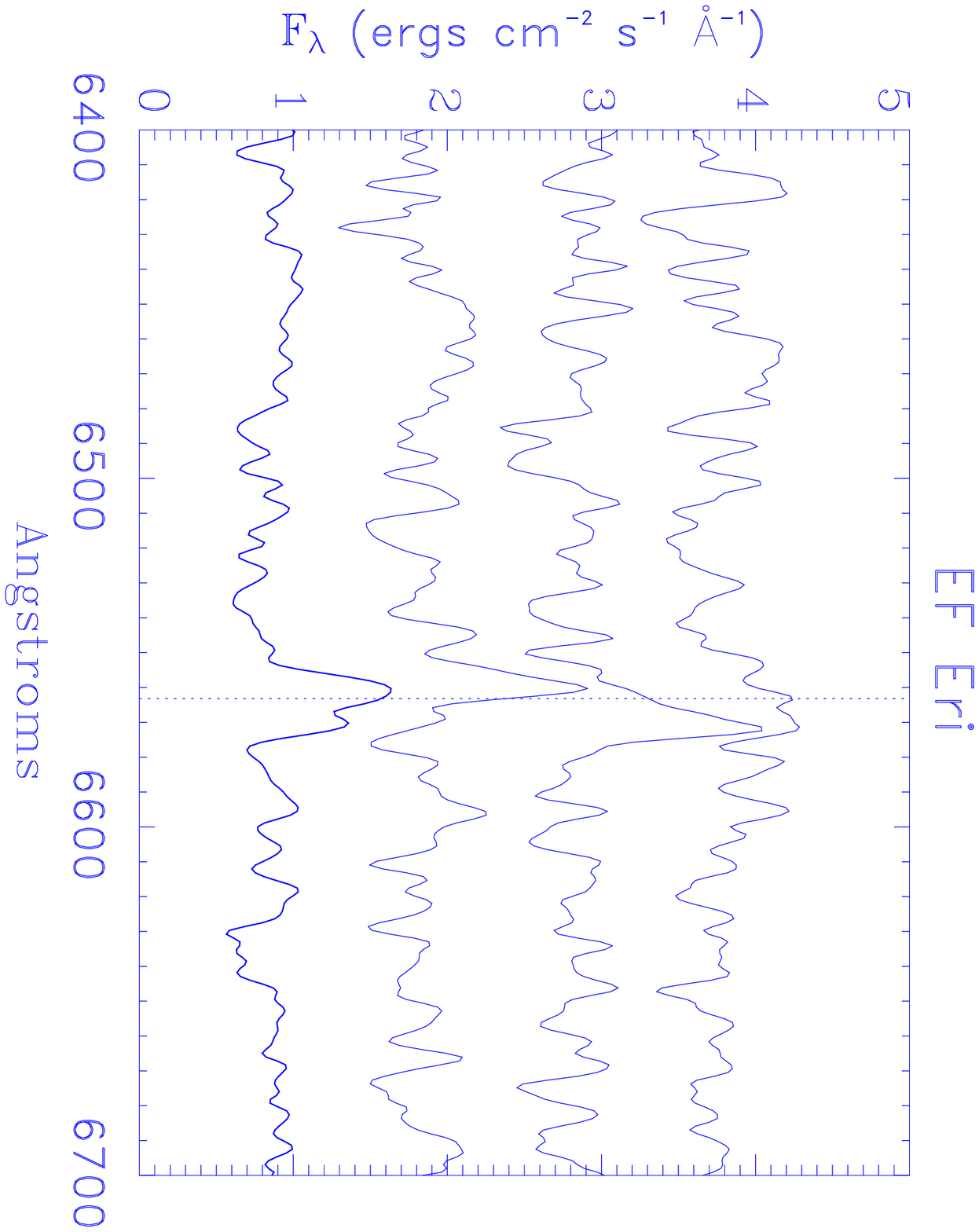}
\end{figure}

\begin{figure}
\plotone{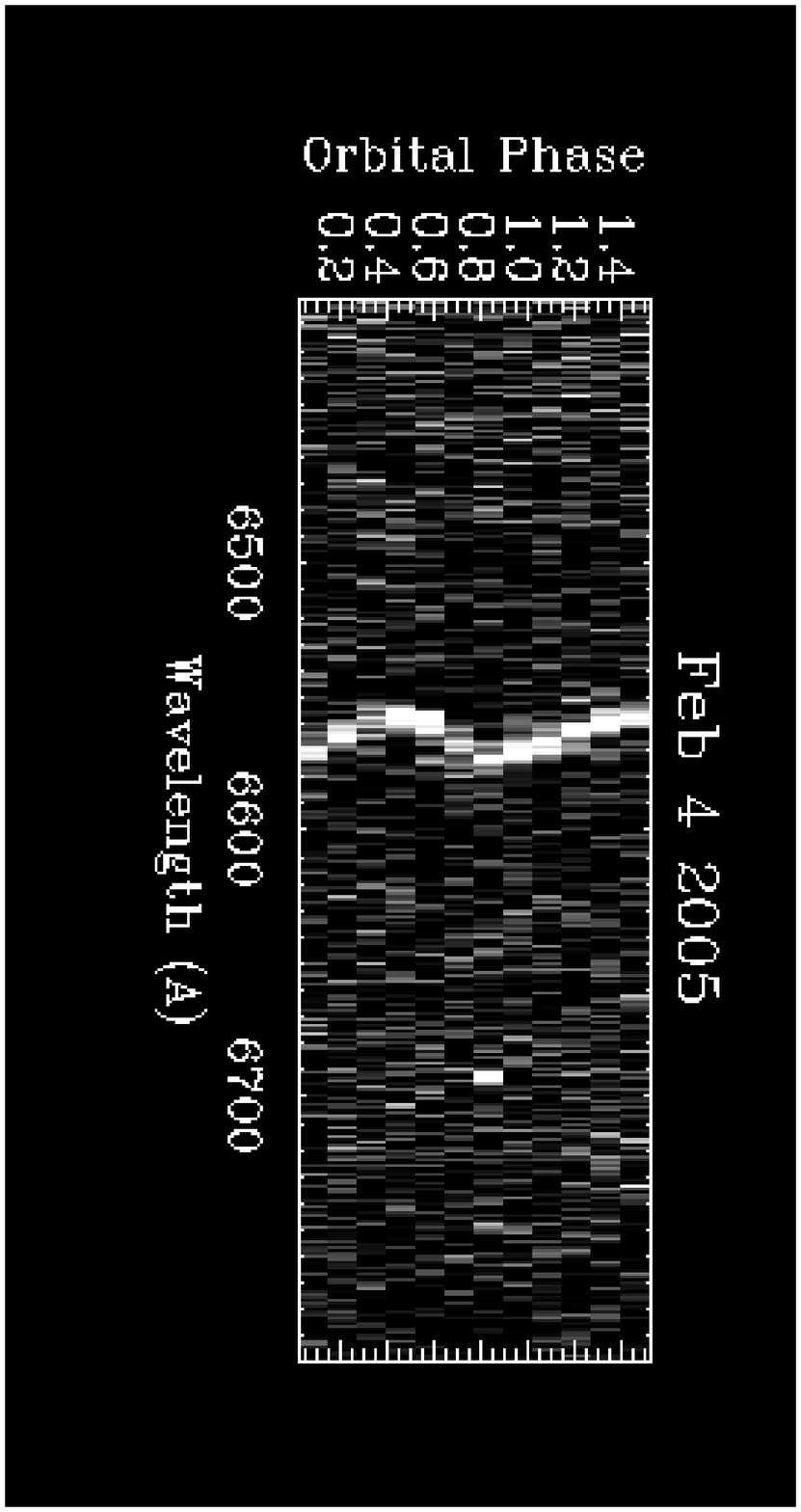}
\end{figure}

%
%
%


\begin{figure}
\plotone{efsmarts_fig6.ps}
\end{figure}

\begin{figure}
\plotone{efsmarts_fig7.ps}
\end{figure}

\begin{figure}
\plotone{efsmarts_fig8.ps}
\end{figure}

\begin{figure}
\plotone{efsmarts_fig9.ps}
\end{figure}

\begin{figure}
\plotone{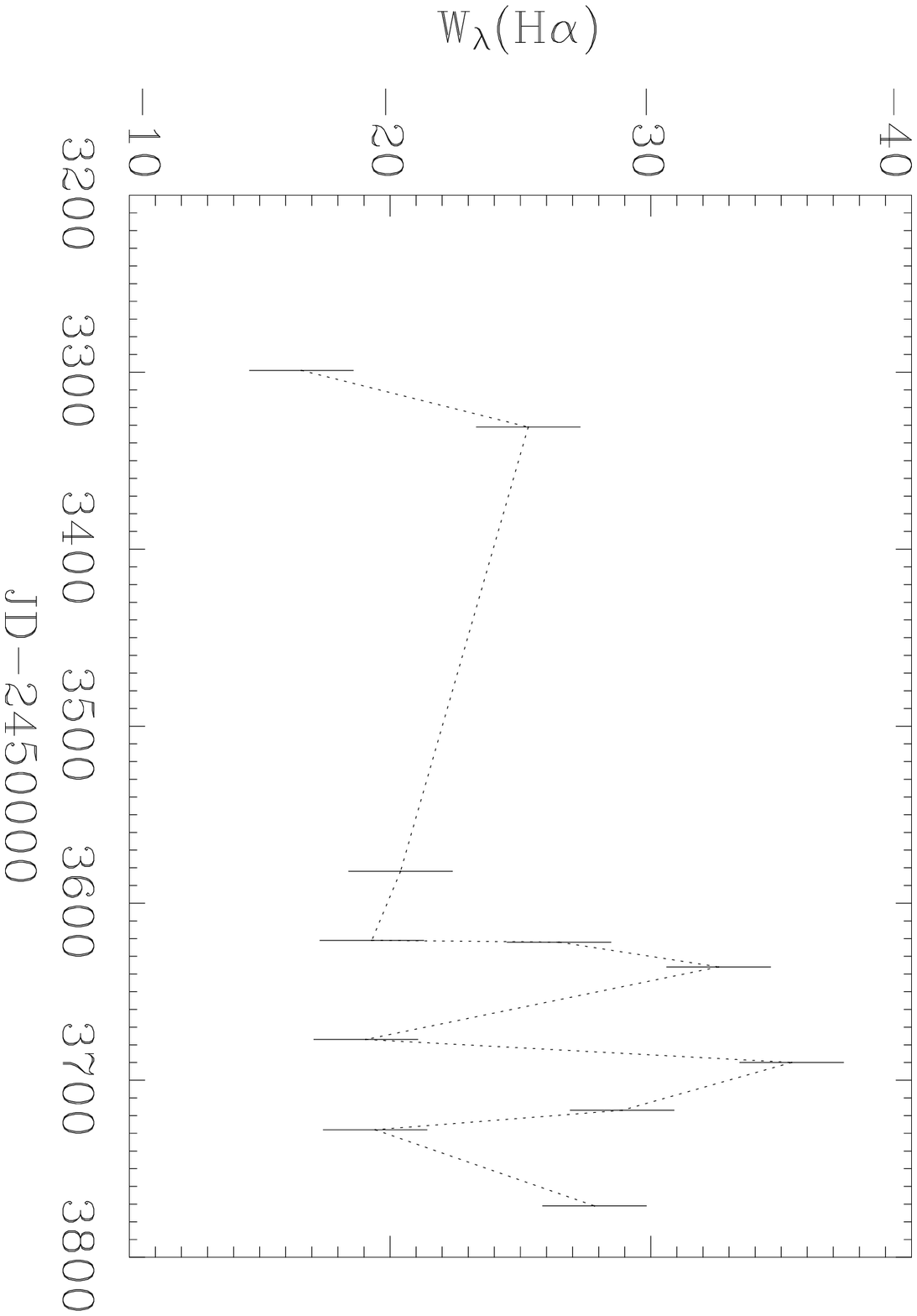}
\end{figure}

\begin{figure}
\plotone{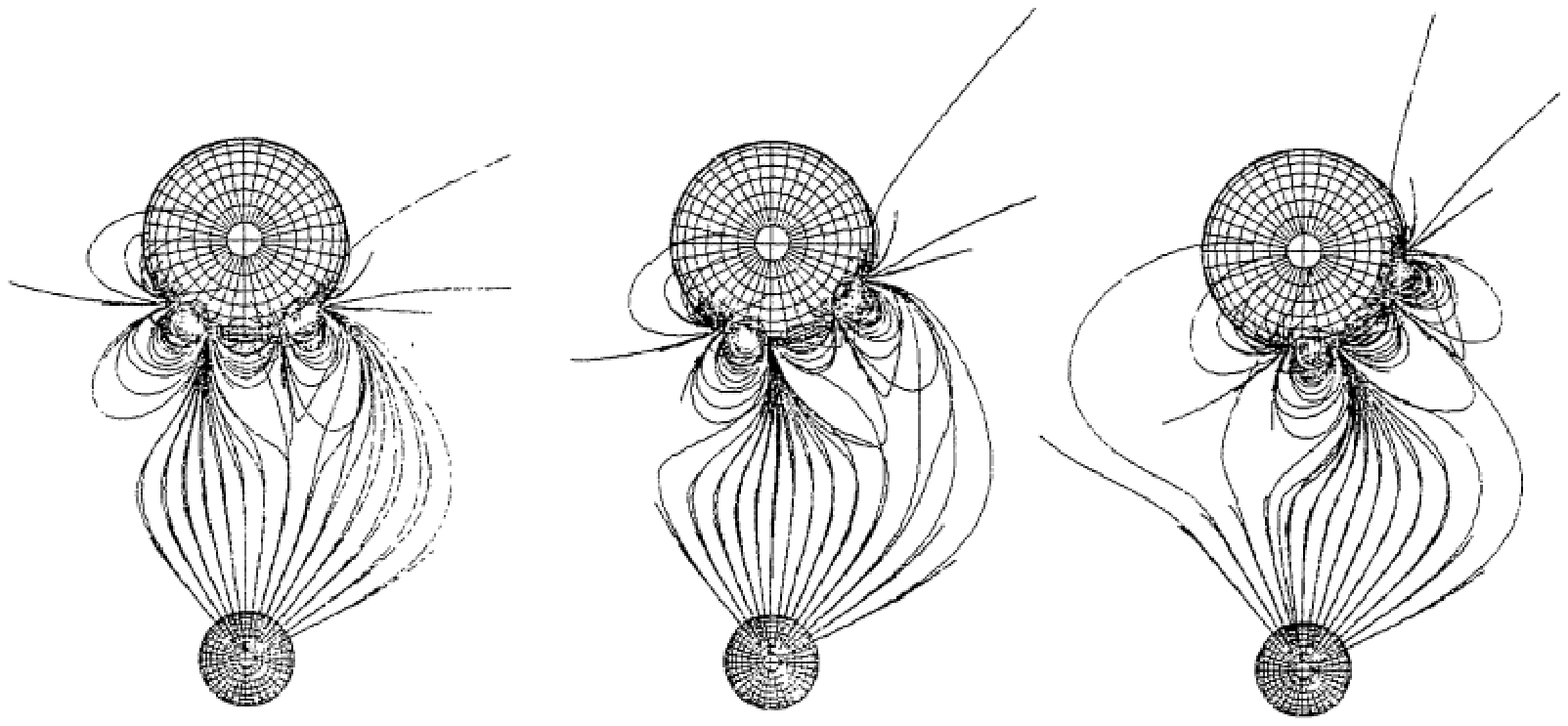}
\end{figure}

\begin{figure}
\plotone{efsmarts_fig12a.ps}
\end{figure}

\begin{figure}
\plotone{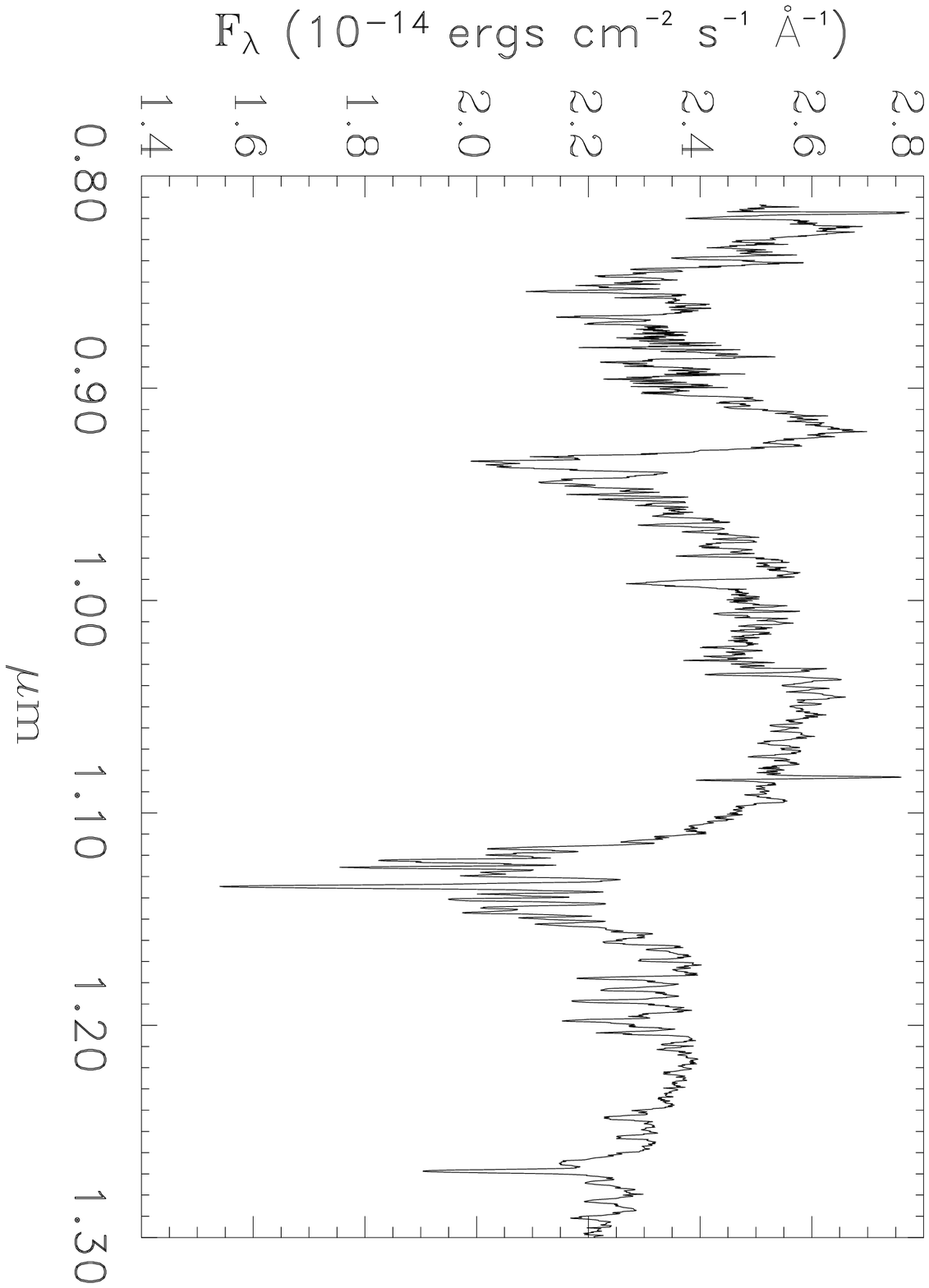}
\end{figure}

\end{document}